# Deep learning for size-agnostic inverse design of random-network 3D printed mechanical metamaterials

H. Pahlavani[a,1], K. Tsifoutis-Kazolis[a], P. Mody[b], J. Zhou[a], M. J. Mirzaali[a], A. A. Zadpoor[a]

[a] *Department of Biomechanical Engineering, Faculty of Mechanical, Maritime, and Materials Engineering, Delft University of Technology (TU Delft), Mekelweg 2, 2628 CD, Delft, The Netherlands*

[b] *Division of Image Processing (LKEB), Radiology, Leiden University Medical Center (LUMC), Albinusdreef 2, 2333 ZA, Leiden, The Netherlands*

**ABSTRACT**

Practical applications of mechanical metamaterials often involve solving inverse problems where the objective is to find the (multiple) microarchitectures that give rise to a given set of properties. The limited resolution of additive manufacturing techniques often requires solving such inverse problems for specific sizes. One should, therefore, find multiple microarchitectural designs that exhibit the desired properties for a specimen with given dimensions. Moreover, the candidate microarchitectures should be resistant to fatigue and fracture, meaning that peak stresses should be minimized as well. Such a multi-objective inverse design problem is formidably difficult to solve but its solution is the key to real-world applications of mechanical metamaterials. Here, we propose a modular approach titled "Deep-DRAM" that combines four decoupled models, including two deep learning models (DLM), a deep generative model (DGM) based on conditional variational autoencoders (CVAE), and direct finite element (FE) simulations. Deep-DRAM (deep learning for the design of random-network metamaterials) integrates these models into a unified framework capable of finding many solutions to the multi-objective inverse design problem posed here. All microarchitectural designs are based on random-network (RN) unit cells with ordered lattices. The first DLM predicts the anisotropic elastic properties of RN unit cells given their microarchitectural design while the other one does the same with the additional input of specimen dimensions. The DGM generates unit cells that give rise to a given set of elastic properties. The integrated framework first introduces the desired elastic properties to the DGM, which returns a set of candidate designs. The candidate designs, together with the target specimen dimensions are then passed to the DLM which predicts their actual elastic properties considering the specimen size. After a filtering step based on the closeness of the actual properties to the desired ones, the last step uses direct FE simulations to identify the designs with the minimum peak stresses. Using an extensive set of simulations as well as experiments performed on 3D printed specimens, we demonstrate that: 1) the predictions of the deep learning models are in agreement with FE simulations and experimental observations, 2) an enlarged envelope of achievable elastic properties (including such rare combinations as double-auxetic behavior and high stiffness) is realized using the proposed approach, and 3) the proposed framework can provide many solutions to the multi-objective inverse design problem posed here.

---

[1] Corresponding author. Tel.: +31-613221032, *e-mail:* h.pahlavani@tudelft.nl

# 1. INTRODUCTION

The second and third decades of the 21$^{st}$ century have witnessed the emergence of architected materials with bespoke, unusual properties that stem from their small-scale design. At the nexus of rational design techniques, where computational models are used to establish design-property relationships, and additive manufacturing (AM, = 3D printing) techniques, which enable the realization of arbitrarily complex designs, a highly vibrant sub-discipline has emerged that is rapidly pushing such designer materials into applications in medical devices[1–3], soft robotics[4–6], and other advanced areas of research[7–9]. Depending on the type of the properties targeted, these architected materials may be referred to as mechanical metamaterials[10–14], acoustic metamaterials[15–17], or meta-biomaterials[3,18], among other types.

Despite their recent academic success, there are two major challenges that hinder the real-world applications of metamaterials in general and mechanical metamaterials in particular. To put these challenges in perspective, let us consider a typical device design scenario where the required elastic properties as well as the dimensions of a device are specified by the device designer. The design problem is then reduced to the problem of finding the microarchitectures that give rise to the required elastic properties while also satisfying the size requirements. The inverse problem of finding the microarchitecture(s) resulting in a specific set of elastic properties is challenging enough in its own right particularly given that the desired combination of properties is often very rare (*e.g.*, high stiffness and highly negative values of the Poisson's ratio [19]). The difficulty of such an inverse design problem is further exacerbated by the fact that most mechanical metamaterials are usually only analyzed in terms of their homogenized or asymptotic properties (*i.e.*, when the number of the constituting unit cells approaches infinity). Such homogenized solutions are only valid at their convergence limits and may significantly deviate from the actual elastic properties when the number of unit cells is not large enough[20]. Given the limited resolution of AM techniques, it is often impossible to fit a very large number of unit cells within a given set of dimensions. Homogenized solutions may, therefore, not offer too much help when dealing with real-world design problems. The inverse design problem should, therefore, not be solved for the asymptotic case of an infinite number of unit cells but for the actual case of a finite number of unit cells in each spatial direction. Here, we use deep learning (DL) models and deep generative models to tackle such a size-agnostic inverse design problem within the context of random-network (RN) mechanical metamaterials.

Most of the mechanical metamaterials developed to date are composed of periodic unit cells. Previous studies have, however, shown that RN units cells, consisting of stretch- and bending-dominated beam-like structures, allow for a wide range of conventional and auxetic elastic properties[21–25], which may go beyond the limits achieved by geometrically-ordered mechanical metamaterials, particularly when seeking after rare combinations of elastic properties[26]. We will, therefore, use RN designs to increase the chance of finding accurate solutions for the inverse design problems targeted here. A facet of such nonlinear inverse problems relevant to microarchitecture design of mechanical metamaterials concerns the non-uniqueness of the solution. It is important to realize that different solutions to the inverse problem posed in the previous paragraph are not equal in many other aspects. That is because designs with similar effective properties could have highly different stress distributions and, thus, highly different degrees of resistance to fatigue and fracture. We are, therefore, interested to find as many of solutions to the posed inverse problem as possible so that additional design requirements, such a uniform stress distribution or a minimum stress peak, can be applied. This further increases the practical utility of the approach presented here.

The existing DL models used for such inverse-design problems are often deterministic in nature. Such models are not well equipped to regress a single input to multiple outputs and may converge to the average of the solutions instead. We will, therefore, model the aforementioned inverse design problem in a probabilistic, generative manner because such approaches have been shown to enable investigations of the structure–response relationship and can resolve the one-to-many mapping problem that deterministic models are unable to cope with[27–30]. Generative adversarial networks (GAN)[31] and variational autoencoders (VAE)[32], which seek to understand the underlying relationship between design features and targets/labels and generate new designs from a low-dimensional latent space, are popular deep generative models used for the inverse design of materials[28,33–36]. In contrast to VAE, which provides a straightforward mapping from the observed dataset to a continuous latent space, a continuous latent space with a meaningful structure is intractable for GAN models[28].

To achieve the goals presented above, we take a modular approach, hereafter referred to as "Deep-DRAM". Deep-DRAM (deep learning for the design of random-network metamaterials) is composed of a sequence of DL and generative models that not only collectively solve the size-agnostic, inverse design problem but can also be (individually) used for many other purposes. First, we create a DL-based forward predictor model that predicts the anisotropic elastic properties of a specific type of RN unit cells. Second, we present a generative model based on conditional variational autoencoder (CVAE) that generates the microarchitecture of RN unit cells with a given set of anisotropic elastic properties. The third module is a DL-based forward predictor model that receives the microarchitecture of the RN unit cells and the desired dimensions of the specimen (*i.e.*, the number of RN unit cells along each spatial direction) and predicts its elastic properties. The developed models are then combined to solve the size-agnostic design problem with the additional requirement that the maximum stresses are minimized. While the data required for training and testing the DL models are all generated using finite element (FE) models, we also present several experiments in which actual mechanical metamaterials are 3D printed and mechanically tested to compare their measured elastic properties and deformation patterns with our computational results.

## 2. RESULTS AND DISCUSSION

### 2.1. Elastic properties of RN unit cells

For the first module, we considered RN unit cells composed of 16 nodes ($n_x \times n_y$, $n_x = n_y = 4$) because this number of nodes allows for a broader range of elastic properties when compared to larger sizes of RN unit cells (Supplementary Figure 2 and Supplementary Table 1), as well as a higher chance of extreme negative and extreme positive Poisson's ratios (Supplementary Table 2). For this number of nodes, it is possible to generate unit cells with average nodal connectivity values of $Z_g = 2.5, 3, 3.5, 4,$ and $4.5$. Depending on the $Z_g$ value, the number of beam-like elements in the RN unit cells varied between 20 and 42 (Figure 1a). It should be noted that the total estimated number of unit cells that can be generated, whether they abide by the design limitations or not, considering the above-mentioned values of $Z_g$ is $\approx 9.22 \times 10^{11}$ (Supplementary Table 3). Assuming that the average simulation time per design equals $\approx 5.42 \times 10^{-4}$ s (Supplementary Figure 3), it takes approximately 497 million seconds ($= 5761$ days) (Supplementary Table 3) to perform FE analysis on all these RN unit cells. The huge number of possible designs highlights the need to have an ultra-fast model to predict the elastic properties of the RN unit cells.

We performed FE analysis on 6 million randomly generated RN unit cells (*i.e.*, 1.2 million unit cells from each group of $Z_g$) as the training group. The elastic properties in directions 1 and 2, which were calculated by FE modeling of these unit cells, cover a cone-like region with a range of $(0, 0.25)$ and $(-1.5, 1.2)$ for the relative elastic moduli ($E_{11}/E_b$ and $E_{22}/E_b$, where $E_b$ is the elastic modulus of the bulk material) and Poisson's ratios ($v_{12}$ and $v_{21}$), respectively (Figure 1b). The distributions of the relative elastic modulus and Poisson's ratio in directions 1 and 2 had similar ranges of values. Moreover, the results show that the RN unit cells are highly anisotropic. The broad range of the elastic properties is due to the possibility to generate both stretching- and bending-dominated structures using random distributions of elements as well as by changing $Z_g$. These results confirm that it is possible to devise RN unit cells with extreme positive and extreme negative values of the Poisson's ratio as well as rare-event[19] double-auxetic unit cells.

We selected six unit cells from the different groups of elastic properties, *i.e.*, almost extreme positive and negative values of the Poisson's ratio in one direction, almost extreme double-auxeticity, almost extreme elastic moduli in both directions, and moderate positive and negative values of the Poisson's ratio as well as moderate values of the elastic moduli in both directions (Figure 1b). To validate the results of our simulations, we 3D printed and experimentally evaluated the elastic properties and deformation patterns of these six unit cells. The experimentally obtained values of the elastic moduli (see the stress-strain curves in Supplementary Figure 4) and Poisson's ratios show a good agreement with FE simulations (Supplementary Table 4). Moreover, the deformation patterns follow similar trends in both simulations and experiments (Figure 1c). In some elements within the FE models, we see higher levels of deformations predicted by FE models as compared to those observed experimentally. These small differences may be explained by the assumptions of the FE models, including a linear elastic constitutive behavior and fully fixed boundary conditions.

We trained a DL model, hereafter referred to as the "unit cell elastic properties model" that predicted the four elastic properties of any RN unit cell given its design (Figure 1d). Based on the results of our hyperparameter tuning pipeline, the applications of an undersampling process and a MinMaxScaler to the cross-validation data resulted in the best model performance. The hyperparameter tuning suggested a model with four hidden layers (500, 376, 252, and 128 hidden neurons in subsequent layers) without a regularization term, with Adam optimizer[37], and with ReLU activation functions throughout the layers (Supplementary Table 5). Within 200 epochs of model training with the optimized hyperparameters, the mean squared error (MSE) and the mean absolute error (MAE) reduced for the training dataset from $9.5 \times 10^{-4}$ and $2.0 \times 10^{-2}$ to $1.5 \times 10^{-5}$ and $2.7 \times 10^{-3}$, respectively. For the validation dataset, the values of MSE and MAE reduced from $3.7 \times 10^{-4}$ and $1.4 \times 10^{-2}$ to $1.6 \times 10^{-5}$ and $2.6 \times 10^{-3}$, respectively (Supplementary Figure 6). The evaluation of the trained model using the test dataset resulted in a coefficient of determination ($R^2$) of >0.993 and >0.999 for the Poisson's ratios and elastic moduli, respectively (Supplementary Table 6 and Supplementary Figure 7). In general, the trained model exhibited a high degree of accuracy in predicting the elastic properties of the RN unit cells with an overall coefficient of determination ($R^2$) of 0.997, an MAE of $3.6 \times 10^{-3}$, and an MSE of $6.0 \times 10^{-5}$ (Supplementary Table 6 and Supplementary Figure 7). All these results show that the model is well trained without underfitting and overfitting and can, therefore, be further used for highly accurate, deterministic prediction of the elastic properties of various RN unit cell designs. The availability of such a model allows for the ultrafast prediction of the elastic properties associated with any design of RN unit cells with the evaluation of the DL model taking $\approx 2.44 \times 10^{-5}$ s per design (for prediction of $10^6$ specimens), which is $> 20$ times faster than the corresponding FE simulation.

**2.2. Generative inverse design framework**

For the inverse design of RN unit cells, we trained a deep generative model based on CVAE that was paired with the pretrained forward predictor (*i.e.*, the unit cell elastic properties model) (Figure 2a). Based on the results of the hyperparameter tuning, the size of the latent space was chosen to be 8. For both recognition and reconstruction models of the CVAE, we selected two hidden layers with 512 and 260 neurons, an Adam optimizer, and ReLU activation functions throughout both hidden layers. ReLU and Sigmoid were selected as the activation functions of the output layer for the recognition and reconstruction models, respectively (Supplementary Table 7 and Supplementary Figure 8).

The reconstruction model of the trained CVAE was separated and called "unit cell generative model". To assess the generative ability of this model, the elastic properties of the test dataset and a random sampling from a normal distribution ($\varepsilon \sim N(0,1)$) were passed as inputs to this model. The predicted unit cell structures were passed as inputs to the unit cell elastic properties model and the predicted elastic properties were compared with the initially requested elastic properties of the test dataset. The results of this comparison showed an overall coefficient of determination ($R^2$) of 0.865, an MAE of $5.1 \times 10^{-2}$, and an MSE of $8.5 \times 10^{-3}$ (Supplementary Figure 9c and Supplementary Table 8). To assess the best achievable accuracy among the designs generated by the unit cell generative model, one hundred possible designs were generated for each set of elastic properties present in the test dataset. Then, the elastic properties of the generated unit cells were compared with the desired mechanical response provided to the model through the calculation of the regression metrics $R^2$, MSE, MAE, and RMSE. The best candidates were then selected among the 100 possible designs. Based on the proposed approach, the final evaluation of the unit cell generative model showed an overall $R^2$ of 0.977, an MAE of $1.2 \times 10^{-2}$, and an MSE of $3.0 \times 10^{-4}$ (Supplementary Figure 9d and Supplementary Table 9). In addition, the high accuracy of the unit cell generative model in generating new RN unit cells was demonstrated by comparing the DL-predicted elastic properties of the generated unit cells with their corresponding FE results (Supplementary Figure 10). Based on this comparison, $R^2$ of 0.98, 0.98, 0.99, and 0.99 were calculated for $\nu_{12}$, $\nu_{21}$, $E_{11}$, and $E_{22}$, respectively (Supplementary Figure 10).

**2.3. Unit cells with requested rare elastic properties**

To demonstrate the generative ability of the unit cell generative model, grid-sampled values of double-auxetic elastic properties, which are rare occurrences in the natural sampling of RNs[21], were created and fed to the deep generative model. For each request, the elastic properties were defined as a combination of $\nu_{12}$ and $\nu_{21}$ in the range of $(-1, -0.1)$ and an equal elastic modulus along both directions with values within the range of $E_{11}/E_b = E_{22}/E_b = (0, 0.25)$. For each input, the best design out of a 100 designs was selected. The four elastic properties of the generated unit cells predicted by the unit cell properties model are reported in a 3D scatterplot incorporating color coding for the fourth property (Figure 2b). To explore the expansion offered by the deep generative model over the observed elastic properties in the initial library (i.e., the training and test datasets), the existing elastic properties in the dataset and the non-duplicate values from the generative process were compared (Figure 2b). From the 3D scatterplot, we can see that the envelope of the achievable elastic properties is expanded in all three planes (i.e., planes of $\nu_{12} - \nu_{21}$, $E_{11}/E_b - \nu_{12}$, and $E_{11}/E_b - \nu_{21}$). The top view of the 3D scatterplot, which shows the expansion of the envelope in the $\nu_{12} - \nu_{21}$ plane, reveals the possibility of generating unit cells with extreme double-auxetic properties (Figure 2b). In general, these results confirm the ability of the deep generative model to generate unit cells with new elastic properties that were not within the envelope of the elastic properties covered by the initial (i.e., training) library. The deep generative model is, therefore, of value for the efficient generation of unit cells with predefined elastic properties, specially rare-event properties, such as double-auxeticity.

As case studies, we selected four sets of elastic properties (I, II, III, and IV) with negative values of the Poisson's ratio and different Young's moduli (see the top views in Figure 2b) to illustrate the generated unit cells corresponding to these cases. For the elastic properties of case IV, three generated candidates are displayed to demonstrate the possibility of generating different designs exhibiting similar sets of elastic properties. The deformation patterns of these unit cells (when subjected to 5% strain along directions 1 and 2) were compared to the initial state of the generated unit cell (Figure 2c). The calculated error values averaged over the four components of the elastic properties were 8.3%, 8.6%, 3.6%, and 3.1% for unit cells I, II, III, and IV, respectively (Supplementary Table 10). These case studies show a high degree of accuracy of the deep generative model when used for the design of double-auxetic unit cells with elastic properties which were not seen before in the training or test datasets.

## 2.4. Elastic properties of combinatorial designs

We studied combinatorial designs composed of $D_1 \times D_2$ repetitions of RN unit cells (Figure 3a). Assuming $D_1$ and $D_2$ are values varying in a range of $(2, 20)$, we studied a total of 100 combinatorial designs from each RN unit cell. We used the undersampled dataset of RN unit cells (dataset size = 81,569), which was used for the training of the unit cell elastic properties model, and performed numerical simulations for all the 100 combinatorial designs composed of these RN unit cells (size of dataset = 8,156,900). The generated dataset was further used to train a forward predictor called "size-agnostic model" that predicts the elastic properties of the combinatorial designs (Figure 3a).

To train the model, we assumed MinMaxScaler as the scaling method, ReLU as the activation function of all the hidden layers as well as of the output layer, and Adam as the optimizer with a learning rate of 0.0001, which were adopted from the hyperparameter tuning step of the unit cell elastic properties model. The hyperparameter tuning step of the deep generative model resulted in 6 hidden layers with 512, 428, 343, 258, 174, and 89 neurons, respectively (Supplementary Table 11). Within 200 epochs, the prediction errors (MSE, MAE) reduced from MSE = $4.7 \times 10^{-3}$ and MAE = $2.9 \times 10^{-2}$ to MSE = $1.6 \times 10^{-5}$ and MAE = $2.9 \times 10^{-3}$ for the training dataset and from MSE = $4.1 \times 10^{-3}$ and MAE = $1.4 \times 10^{-2}$ to MSE = $1.7 \times 10^{-5}$ and MAE = $3.0 \times 10^{-3}$ for the validation dataset (Supplementary Figure 11). The trained model had an overall $R^2$ of 0.995 for the test dataset (10% of the original dataset), confirming that it can accurately predict the elastic properties of the combinatorial designs (Supplementary Table 12 and Supplementary Figure 12).

Combinatorial designs showed a wide range of elastic properties. The relative elastic moduli ($E_{11}/E_b$ and $E_{22}/E_b$) and Poisson's ratios ($\nu_{12}$ and $\nu_{21}$) calculated by the numerical simulations were in the ranges of $(0, 0.3)$ and $(-2, 3)$, respectively (Figure 3b). The 3D distribution of the elastic properties of the combinatorial designs resembled a square pyramid with inwardly curved faces whose base is placed within the $\nu_{12} - \nu_{21}$ plane. The distribution of the elastic properties in the $\nu_{12} - \nu_{21}$ plane was bounded by two hyperbolas, one with openings in the first and third quadrants and the other one with openings in the second and forth quadrants.

To study how the elastic properties vary with $D_1$ and $D_2$, we selected one of the RN unit cells and depicted the evolution of the four elastic properties as a function of changes in $D_1$ and $D_2$ (Figure 3c). We found a nonlinear relationship between the elastic properties and dimensions of the combinatorial designs of this unit cell that, as expected, saturates for large enough numbers of unit cells along each spatial direction (Figure 3c and Supplementary Figure 13). For this selected RN unit cell, $E_{11}/E_b$, $E_{22}/E_b$, $\nu_{12}$, and $\nu_{21}$ converge towards 0.020, 0.048, 0.05, and 0.5, respectively (Supplementary Figure 13). Based on a preliminary study we performed on the combinations of RN unit cells, we selected $D_1 = D_2 = 20$ as the maximum size of combinatorial designs due to the saturation of the elastic properties for larger numbers of unit cells (Supplementary Figure 13). To validate the results of our simulations, the elastic properties and deformation patterns of four selected combinatorial designs were determined experimentally (Figure 3d). The mechanical tests on these specimens indicated that the deformation patterns follow the same trends as observed in the simulations. The mismatches between the simulations and mechanical tests can be attributed to the assumptions used in the simulations (*e.g.*, a linear elastic constitutive equation), differences between the experimental and simulated boundary conditions, and manufacturing imperfections.

**2.5. Inverse design of lattice structures with requested elastic properties and dimensions**

We combined the unit cell generative model and the size-agnostic model to develop a comprehensive and powerful framework called Deep-DRAM, which can inversely design lattice structures their given elastic properties and dimensions. Given the requested elastic properties, we first used the deep generative model to generate $10^5$ RN unit cells. It takes the deep generative model $5.7 \pm 0.1$ s on a workstation (see the Methods section for the specifications) to generate these unit cells. Since the returned unit cell structures would vary in their design and mechanical response, they can deviate from the requested set of elastic properties. This is, in fact, an advantage of such a generative model because the actual elastic properties of a lattice structure with a finite (anisotropic) number of unit cells along each spatial direction may be quite different from the unit cell properties. The presence of such natural variations in the elastic properties of the generated unit cells enables us to feed a large number of designs created by the deep generative model to the size-agnostic model that is trained to account for the effects of size along each direction. Then, all these generated unit cells together with the desired dimensions are introduced to the size-agnostic model to predict the elastic properties of these combinatorial designs. Finally, the MSE values showing the difference between the target properties and the final elastic properties of the generated combinatorial designs are calculated. Based on the error values, we selected the designs that best matched the target elastic properties for the given dimensions (Figure 4a).

To demonstrate the functionality of Deep-DRAM, we assumed a constant value for the elastic modulus (*i.e.*, $E_{11}/E_b = E_{22}/E_b = 0.03$) and a range of $(-1, 1.6)$ for the Poisson's ratios with a step size of $0.2$ (*i.e.*, $14$ groups of values for the Poisson's ratios). In total, we studied $196$ (*i.e.*, $14 \times 14$) sets of elastic properties. We also predefined a dimension of $D_1 = D_2 = 4$ for the generated combinatorial designs. Using these predefined values, $1.96 \times 10^7$ combinatorial designs were generated and filtered based on their MSE values. The whole design procedure including the inverse design of the RN unit cells, combining the unit cells into combinatorial designs, prediction of the elastic properties of the combinatorial designs, and finding the best candidates based on the calculated MSE values took ≈38 min (for all the $1.96 \times 10^7$ designs) using the same, above-described computer.

To quantify the expected error values for the design of combinatorial designs with different elastic properties and dimensions, we repeated the aforementioned procedure for 196 selected sets of elastic properties considering two groups of dimensions. In the first group, we predefined equal dimensions (*i.e.*, $D_1 = D_2 = [2, 4, 6, 8, 10, 12, 14, 16, 18, 20]$) while in the second group we assumed $D_2 = 2$ and $D_1 = [2, 4, 6, 8, 10, 12, 14, 16, 18, 20]$. We defined the envelope of successful designs such that it was bounded by the designs corresponding to an MSE value of $0.1$. The heat maps of the MSE values depict the expected error values for generating combinatorial designs when the elastic properties and dimensions are provided as input (Figure 4b). The gray regions represent the designs whose elastic properties are associated with MSE values exceeding the acceptance threshold (Figure 4b). Upon closer inspection, we found that the gray regions primarily correspond to the property-size combinations that simply cannot arise from the considered RN. As expected, the envelopes of successful design generations converge for large enough values of $D_1$ and $D_2$ (Figure 4b). For larger sizes, the heat maps of MSE values for combinatorial designs with $D_1 = D_2$ are symmetrical around the $v_{12} = v_{21}$ line. This increased symmetry indicates a more isotropic behavior of RN combinatorial designs as their dimensions increase. In the other group of the combinatorial designs with $D_2 = 2$ and varying $D_1$ values, we do not expect isotropy because the ratio of $D_1/D_2$ increases and the geometry is not symmetric anymore.

## 2.6. Stress distribution

Deep-DRAM provides many solutions to the design problem of finding RN lattice structures with pre-defined dimensions and elastic properties. It is, therefore, possible to apply additional design requirements, such as criteria regarding the stress distributions observed within the generated structures under various types of loading conditions. One such criterion is to choose the design with the minimum peak stress, thereby enhancing their resistance against fatigue and failure. To demonstrate the utility of our size-agnostic inverse design framework within this context, we first generated combinatorial designs with predefined elastic properties and dimensions. We then filtered the generated designs based on their maximum von Mises stress (Figure 5a). As representative cases, we studied three groups of combinatorial designs with predefined specifications: i) $D_1 = D_2 = 4$, $v_{12} = -0.2$, and $v_{21} = 0.2$, ii) $D_1 = 10$, $D_2 = 4$, $v_{12} = v_{21} = 0.5$, and iii) $D_1 = D_2 = 10$, $v_{12} = v_{21} = -0.2$, while the elastic modulus was assumed to be the same for these three groups (*i.e.*, $E_{11}/E_b = E_{22}/E_b = 0.03$). From each group, the first 1000 designs with MSE < 0.1 were further analyzed using FE simulations to determine the stress distribution within their elements under two loading conditions (*i.e.*, $\varepsilon_{11} = 5\%$ or $\varepsilon_{22} = 5\%$). The normalized peak values of the von Mises stress in directions 1 and 2 were then calculated (Figure 5b). From each group, we selected two specimens with almost the same MSE but with either the minimum or maximum Euclidean distance from the origin. The stress distributions corresponding to these case studies clearly show stress concentrations in some regions within the specimens with the maximum Euclidean distances (*i.e.*, specimens 2, 4, and 6) while the stress distribution are comparatively more uniform within specimens 1, 3, and 5 (Figure 5c). The specimens with high peak stresses are prone to premature crack initiation and growth and should be avoided in the design of mechanical metamaterials aimed for practical applications. Further analysis of these results shows 310%, 250%, and 270% difference between the maximum and minimum values of the von Mises stresses of the three study groups, which are very substantial numbers within the context of peak stress reduction analysis.

## 3. DISCUSSIONS AND FUTURE OUTLOOK

The Deep-DRAM framework presented here is a combination of four modules and provides many opportunities for the design of mechanical metamaterials for practical use in the design of advanced functional devices. In addition, the presented modular approach allows the individual modules to be combined with other tools available elsewhere to provide solutions for the many challenges encountered in the design of designer materials. To a degree, the modularity of this approach and the probabilistic nature of the CVAE allows us to decouple some of the problems encountered in the design and optimization of mechanical metamaterials, thereby enabling multi-objective design optimization with minimum development and computational costs. For example, the multiple objectives of achieving a certain set of elastic properties and minimizing the peak stress within the structure can be handled in-series with minimum computational costs. That is partially due to the extremely high speeds of both generating and evaluating individual designs, which are in the range of micro-seconds.

There are a number of points that need to be discussed regarding the broader use of Deep-DRAM. First, while we focused on a specific choice of RN for this study, the same methodology can also be used for any underlying design paradigm including any other types (*i.e.*, size, organization) of random structures as well as ordered structures and a combination thereof. Second, the modular design of our approach as well as its *ad hoc* combination with direct FE modeling affords it a high degree of flexibility in terms of taking design requirements into account and tackling multiple types of problems that are challenging in their own right. For example, the problem of finding rare combinations of elastic properties is treated independently in multiple other studies[19,38] but can also be studied, within the confines of the selected RN design, using the modules developed here. Third, our focus on the linear elastic properties meant that we used linear elastic constitutive models everywhere in the current study. However, the same approach can be used to study the nonlinear properties of RN designs or to consider any other aspects of their constitutive behavior (*e.g.*, viscoelasticity). The only difference would be that the FE models need to be modified to reflect the more complex constitutive behavior. Indeed, the relative advantage of the presented approach would be even more evident when the simulation time is longer, such as the case of nonlinear or viscoelastic constitutive behaviors. Fourth, the compact and computationally efficient nature of the final models means that they can be implemented in low-resource settings to power edge computing[39,40] applications. Finally, some elements of the developed modulus (*i.e.*, even individual layers) can be used for more advanced machine learning approaches, such as transfer learning, to further generalize the domain of application of our models.

## 4. CONCLUSIONS

In summary, we developed a size-agnostic inverse design framework, Deep-DRAM, which can generate RN lattice structures not only with predefined elastic properties but also with predefined dimensions suitable for any intended application. We showed that combining deep generative models with forward predictors is successful in generating bespoke mechanical metamaterials while also satisfying additional design requirements, such as minimum peak stresses, to improve the endurance of designer materials for real-world applications.

## 5. METHODS

We studied restricted RN unit cells in which the nodal points of the beam-like elements were fixed at specific locations. The design of these unit cells was inspired by our previous research[21] that computationally explored the auxeticity and stiffness of RNs and demonstrated a wide range of elastic moduli and Poisson's ratios for this type of mechanical metamaterials. For this work, we first studied RN unit cells composed of different node numbers (*i.e.*, $n_x = n_y = 3, 4, 5, 6, 7,$ and $8$) to find out the least number of nodes that corresponds to the broadest range of elastic properties. We assumed the internodal distances of $\Delta x$ and $\Delta y$ in directions 1 and 2, respectively. The overall size of a unit cell is, therefore, given by: $L \times W$ ($L = n_y \times \Delta y$ and $W = n_x \times \Delta x$) (Figure 1a). Also we assumed the in-plane ($t = 1$ mm) and out-of-plane ($T = 10$ mm) thicknesses for the RN unit cells. The beam-like elements were randomly distributed to connect the whole grid. We studied unit cells with the network connectivity values of $Z_g = 2.5, 3, 3.5, 4,$ and $4.5$ (Figure 1a). We further studied combinatorial designs that are composed of different numbers of rows and columns of RN unit cells.

### 5.1. Computational models

All the FE models were created using MATLAB (MATLAB R2018b, MathWorks, USA) codes. Custom codes were used to design the structures by randomly connecting each node to its surrounding nodes and to perform the FE simulations that estimate the elastic properties of the resulting structures (*i.e.,* the elastic moduli and Poisson's ratios in both orthogonal in-plane directions). The random distribution of beams resulted in 'loose designs' where some nodes were not connected to the overall grid. To exclude such designs, we used a graph-based search method (breadth-first search[41]) for filtering and discarding such invalid unit cell designs. The applied graph-based algorithm sped up the process by nearly 900 times as compared to an image-based filtering method used previously[21] (more information is provided in Supplementary Figure 1). Our codes were further extended to incorporate the RN unit cells into the combinatorial designs (Figure 3a).

We employed three-node quadratic beam elements (Timoshenko beam elements) with rectangular cross-sections ($t \times T$) and with two translational (*i.e.*, $u_{11}$ and $u_{22}$) and one rotational (*i.e.*, $u_{33}$) degrees of freedom (DOF) at each node. An elastic material with a Young's modulus of $E_b = 0.6$ MPa and a Poisson's ratio of $\nu_b = 0.3$ was then assigned to elements. For each structure, two FE models were created to separately apply a strain of 5% along 1- and 2-directions. In the first model, the top nodes were subjected to a strain of 5% along the 2-direction ($u_{11} = u_{33} = 0$ and $u_{22} = 0.05 \times L$) while all the DOF of the bottom nodes were constrained ($u_{11} = u_{22} = u_{33} = 0$). In the second model, the right nodes were subjected to 5% strain along the 1-direction ($u_{22} = u_{33} = 0$ and $u_{11} = 0.05 \times W$) while all the DOF of the left nodes were constrained ($u_{11} = u_{22} = u_{33} = 0$). More information about the FE equations used for the numerical simulations are provided in the supplementary document.

To calculate the elastic moduli of the structures ($E_{11} = \sigma_{11}/\varepsilon_{11}$ and $E_{22} = \sigma_{22}/\varepsilon_{22}$), the normal stresses along directions 1 and 2 ($\sigma_{11} = \overline{F_{11}}/(L \times T)$, $\sigma_{22} = \overline{F_2}/(W \times T)$ (Figure 1a)) were divided by the strain applied along the same direction ($\varepsilon_{11} = \varepsilon_{22} = 5\%$). In these equations, $\overline{F}_{11}$ and $\overline{F}_{22}$ are the mean reaction forces along directions 1 and 2 at the right and top nodes, respectively ($\overline{F_{11}} = \frac{\sum_{i=1}^{n_R} F_{11,i}}{n_R}$, $\overline{F_{22}} = \frac{\sum_{i=1}^{n_T} F_{22,i}}{n_T}$, where $n_R$ and $n_T$ are the total numbers of the right and top nodes while $F_{11,i}$ and $F_{22,i}$ are the reaction forces along directions 1 and 2 at each of the right and top nodes, respectively). We then calculated the transverse strain as the ratio of the average displacement of the lateral nodes to the initial transversal length of the structure (in the case of $\varepsilon_{axial} = \varepsilon_{11} = 5\%$: $\varepsilon_{trans} = \varepsilon_{22} = \frac{\sum_{i=1}^{n_T} \delta y_i}{L \times n_T}$, and in the case of $\varepsilon_{axial} = \varepsilon_{22} = 5\%$: $\varepsilon_{trans} = \varepsilon_{11} = \frac{\sum_{i=1}^{n_R} \delta x_i}{W \times n_R}$). The transverse strain was then divided by the applied axial strain to calculate the Poisson's ratio ($\nu = -\frac{\varepsilon_{trans}}{\varepsilon_{axial}}$).

### 5.2. Deep learning

**Unit cell elastic properties model**: We trained a predictor model that we refer to as the "unit cell elastic properties model" which aims to learn the mapping from the space of RN unit cell designs to that of their elastic properties. This model takes as input a binary vector representing the RN unit cells (*i.e.*, a binary vector of 0 and 1 values, where 1 indicates the presence of an element and 0 indicates its absence) and returns the elastic properties ($E_{11}$, $E_{22}$, $\nu_{12}$, and $\nu_{21}$) of the unit cells as output. Before training the model, we performed an initial data analysis process followed by a hyperparameter tuning study. To select the options and parameters for both data analysis and hyperparameter tuning, we used a pipeline training technique (Supplementary Figure 5) which combined data analysis options and model hyperparameters in its search space. Pipeline training automates the training process including data analysis and hyperparamater tuning and optimizes the model considering different configurations of the parameters of the search space of the pipeline.

We used a workstation (CPU = Intel® Xeon® W-2295, RAM = 256 GB) and a Python script (Python 3.9.7) to tune the parameters of the pipeline's search space. Though this, 10,368 combinations of parameters were investigated (Supplementary Table 5) and the best pipeline parameters were selected. For data analysis, we selected data resampling and data scaling to be investigated since the size and distribution of both inputs and outputs are important for the success of the model training step. As for the model hyperparamaters, we selected the parameters describing the design of the DL models (*i.e.*, the width and depth of the hidden layers as well as the trend of the variation of the number of hidden neurons per layer), the regularization terms, the type of the optimizer algorithm, the activation functions of the hidden layers and output layer, and the application of batch normalization. We used the search methods of cross-validation grid search from scikit-learn (version 1.1.1) to systematically iterate over the predefined values of the search space parameters (see the supplementary document for a more in-depth discussion of the methods).

The overall performance of the model was assessed by characterizing its ability to generalize from the training dataset to the test dataset to avoid both under- and overfitting. To avoid overfitting, we used *k*-fold cross validation (CV) that divides the training dataset into *k* smaller sets. We used 3-fold CV, meaning that each set equals 33% of the training dataset. Note that 10% of the overall dataset was kept as the test dataset for final model evaluation.

We selected MSE (Equation (1)) as the loss function of the model to ensure the regression line changes only slightly for a modest change in the data point. For the evaluation of the model training, we used the coefficient of determination ($R^2$) (Equation (2)) which indicates the amount of target variance explained by the model's independent variables.

$$\text{MSE} = \frac{1}{n}\sum_{i=1}^{n}(y_i - \hat{y}_i)^2, \tag{1}$$

$$R^2 = 1 - \frac{\sum_{i=1}^{n}(y_i - \hat{y}_i)^2}{\sum_{i=1}^{n}(y_i - \bar{y})^2}, \tag{2}$$

where n is the size of the dataset, $y_i$ is the i[th] real target, $\hat{y}_i$ is the corresponding predicted value, and $\bar{y}$ is the mean value of $y$ ($\bar{y} = \frac{1}{n}\sum_{i=1}^{n} y_i$).

**Deep generative model:** We trained a deep generative model that allows for the inverse design of RN unit cells. This model is based on CVAE and follows a similar approach as in a number of previous studies[27,42]. The key difference between CVAE and VAE is that CVAE can incorporate certain conditions in the training process[42,43]. Here, the additional conditions concern the elastic properties of the RN unit cells. Two deep neural networks were utilized as the sub-models of the CVAE, each with a structure purposefully built for their specialized roles. More specifically, they are a recognition network and a reconstruction network which are coupled in an encoder-decoder format (Figure 2a). The recognition model transfers the designs of the RN unit cells as well as their corresponding elastic properties into a low-dimensional, continuous, and ordered latent space[28]. The reconstruction model uses the four elastic properties and the latent variables to recreate the binary vector representing the metamaterial design. After the successful training of the CVAE, the reconstruction model was separated and used as the deep generative model. The input to this deep generative model was the desired elastic properties of the RN unit cell as well as a random sampling from a normal distribution with the same dimensions as the latent space.

The loss function that we utilized to train the CVAE ($\mathcal{L}_{CVAE}$) was obtained from the loss function of a standard VAE ($\mathcal{L}_{VAE}$) with conditional information included. The loss function of VAE consists of two terms, the reconstruction error and the Kullback-Leibler (KL) term, and is given as[32,42,43]:

$$\mathcal{L}_{VAE} = E[logP(x|z)] - D_{KL}[Q(z|x) \| P(z)], \tag{3}$$

where $E$ represents an expectation value, $P$ and $Q$ are probability distributions, $D_{KL}$ represents the Kullback-Leibler divergence, $x$ is the binary vector representing the RN unit cells, and $z$ represents the latent variables. $Q(z|x)$ and $P(x|z)$ are approximated by the recognition and reconstruction models, respectively. The incorporation of the conditional information in the loss function of the VAE modifies the loss function of CVAE as follows[42,43]:

$$\mathcal{L}_{CVAE} = E[logP(x|z,y)] - D_{KL}[Q(z|x,y) \| P(z|y)], \tag{4}$$

where $y$ is a condition vector that plays an active role in both the encoding and decoding operations. The condition vector in our model contains the elastic properties of the RN unit cells.

To assess the elastic properties of the generated RN unit cells and as a regularisation term to the overall loss function[30], the pre-trained "unit cell elastic properties model" that predicts the elastic properties of the RN unit cells was incorporated in the overall loss function ($\mathcal{L}_{all}$). $\mathcal{L}_{all}$ for the training of the deep generative model contains terms that account for the configuration of the latent space, the reconstruction of the input metamaterial design, and the retrieval of the desired elastic properties from the generated RN unit cells. For the retrieval of the desired mechanical properties from the reconstructed RN unit cells, MSE is considered as the loss function, $\mathcal{L}_{MSE}$. Therefore, the total loss function for the deep generative model (Equation (5)) include those corresponding to the CVAE ($\mathcal{L}_{CVAE}$) and the MSE ($\mathcal{L}_{MSE}$) between the target and predicted elastic properties of the reconstructed RN unit cells.

$$\mathcal{L}_{all} = \mathcal{L}_{CVAE} + \mathcal{L}_{MSE} \tag{5}$$

To train a CVAE with an optimal fitting and a lower dimension of the latent space, we used the same hyperparamater tuning pipeline as for the unit cell elastic properties model. Some assumptions were made based on the best resulting parameters for the unit cell properties model. Identical hyperparameters as for the unit cell elastic properties model were used in the data processing steps. Additionally, the Adam optimizer was preselected, having outperformed the RMSprop optimizer in the training of the "unit cell properties model". In total, 6144 combinations of parameters were tested through the hyperparameter optimization method with a running time of approximately 1341 minutes (~23 hours) (Supplementary Table 7). For the validation of the elastic properties arising from the generated RN unit cells and as a regularization term to the overall loss, the pre-trained unit cell elastic properties model was used as a forward predictor of the elastic properties of the RN unit cells ($E_{11}$, $E_{22}$, $\nu_{12}$, and $\nu_{21}$).

For the evaluation of the trained CVAE (Supplementary Figure 8) with the help of the test dataset, we visualized the latent space to see if it is well-clustered. Moreover, we used the relevant performance metrics (*i.e.*, Confusion matrix, Precision, Recall, and $F_1$ score) for the same purpose. To visualize the latent space (Supplementary Figure 9a) where the metamaterial design is encoded, the *t*-distributed stochastic neighbor embedding (*t*-SNE) approach was used to reduce its dimension to two. In addition to capturing the relevant information regarding the design of RN structures, the latent variables need to capture some information regarding the elastic properties of the designs. As a result, elastic properties are examined and distributed inside each geometric cluster. Because elastic properties are continuous and, thus, cannot be split into categories, they were manually classified according to certain specific criteria to investigate if the latent space can identify distinct Poisson's ratios to a satisfactory degree. The targets of the elastic properties are assigned to three classes: 0 for auxetic metamaterials, 1 for conventional metamaterials, and 2 for double-auxetic metamaterials. The binary multilabel class output is assessed in the case of unit cell representation reconstruction. This is accomplished by calculating the weighted average of the actual and predicted classes for each sample in the test dataset. The confusion matrix shows, in a sample-wise manner, the summary of the prediction results of the classification problem with each row corresponding to the actual class and each column corresponding to the predicted one. This matrix was then used to assess the classification accuracy of the model.

For the evaluation of the model training, the $F_1$ score (Supplementary Figure 9b) was chosen as the evaluation index of the introduced RN reconstruction, and the predicted elastic responses of the returned structures were evaluated using $R^2$. The $F_1$ score is mainly established for binary classification tasks with the values 1 and 0 corresponding to the best and worst performances, respectively. The $F_1$ score may be considered a weighted harmonic mean of the precision and recall, where the recall and precision are both equally essential (Equation (6)). Intuitively, precision is the proportion of the true positive cases among those labelled as positive by the model and recall is the proportion of the positive instances among the total number of positive examples including :

$$F_1 = \frac{2 \times \text{recall} \times \text{precision}}{\text{recall} + \text{precision}} \text{, where recall} = \frac{\text{TP}}{\text{TP+FN}}, \text{precision} = \frac{\text{TP}}{\text{TP+FP}}, \tag{6}$$

where TP is true positive, FN is false negative, and FP is false positive.

**Size-agnostic model:** To predict the elastic properties of the combinatorial designs composed of RN unit cells with given dimensions, we trained a forward predictor model referred to as the "size-agnostic model". This model aims to learn the mapping from the space of combinatorial designs composed of $D_1 \times D_2$ repetitions of RN unit cells (where $D_1$ and $D_2$ are the number of the repetitions of a RN unit cell along directions 1 and 2, respectively, and are assumed to be an even value in the range of $(2-20)$) to the space of their elastic properties (i.e., $E_{11}$, $E_{22}$, $\nu_{12}$, and $\nu_{21}$). A binary vector representing the RN unit cells combined with the vectors $D_1$ and $D_2$ were introduced to the model as inputs. The model returned the elastic properties of the combinatorial design as its output. Before training the model, we used the same hyperparameter tuning pipeline as described above for the unit cell elastic properties model to optimize the hyperparamaters of the model (Supplementary Table 11). We assumed MSE (Equation (1)) as the loss function of the model and used $R^2$ (Equation (2)) and MSE (Equation (1)) for the evaluation of the trained model.

## 5.3. Experiments

We selected six RN unit cells (Figure 1b and 1c) and four combinatorial designs (Figure 3d) to be 3D printed and mechanically tested. We used selective laser sintering (SLS) for printing these lattices using a commercially available material (*i.e.*, Oceanz Flexible TPU). We attached the 3D printed specimens to the testing machine using a designed pin and gripper system which was 3D printed using a fused deposition modeling (FDM) 3D printer (Ultimaker 2+, Geldermalsen, the Netherlands) from polylactic acid (PLA) filaments (MakerPoint PLA, 750 gr, Natural). We used a mechanical testing machine (LLOYD instrument LR5K, load cell = 100 N) to perform axial tensile loading test on the specimens (stroke rate = 1 mm/min) along directions 1 and 2. The stress-strain curves were then obtained based on the applied displacements and the recorded reaction forces. Stress and strain values were calculated by dividing the force by the initial cross-section area and dividing the crosshead displacements by the initial length of the specimen, respectively. The overall stiffness of the specimens along directions 1 and 2 (*i.e.*, $E_{11}$ and $E_{22}$) were then calculated as the slope of the stress-strain curves. To calculate the Poisson's ratios (*i.e.*, $\nu_{12}$ and $\nu_{21}$) of the specimens, we performed image analysis using a custom-made MATLAB code. For this purpose, we used a digital camera to capture the lateral deformations of the specimens to measure the transverse strain at the different steps of the applied longitudinal displacement. Finally, the Poisson's ratio was calculated as $\nu = -\frac{\varepsilon_{trans}}{\varepsilon_{axial}}$, where $\varepsilon_{trans}$ and $\varepsilon_{axial}$ were calculated in the same way as in the computational models.

**COMPETING INTERESTS**

The authors declare no competing interests.

## FIGURE CAPTIONS

**Figure 1.** A schematic illustration and elastic properties of the RN unit cells as well as the network architecture of the unit cell elastic properties model. (a) To design the RN unit cells, we predefined the node coordinates with a fixed horizontal and vertical distance of $\Delta x = \Delta y = 7.5$ mm. Assuming a grid of $4 \times 4$ nodes, the overall dimensions of each unit cell is $L = W = 22.5$ mm. Based on the defined overall connectivity, $Z_g$, the applicable number of beam-like elements were randomly distributed within the structure. (b) The elastic properties (*i.e.*, $E_{11}/E_b$, $E_{22}/E_b$, $\nu_{12}$, and $\nu_{12}$) calculated for the RN unit cells using FE analysis. (c) The deformation patterns of six RN unit cells under two loading conditions of $\varepsilon_{11} = 5\%$ and $\varepsilon_{22} = 5\%$ as predicted by FE analysis and observed in the mechanical tests on the 3D printed specimens. (d) The network architecture of the trained unit cell elastic properties model, which maps the design of the unit cell to their elastic properties.

**Figure 2.** The unit cell generative model and its ability to generate new RN unit cells. (a) The CVAE is composed of two parts: 1) the recognition part, which maps the unit cell design and its corresponding elastic properties to the latent space and 2) the reconstruction part, which converts a sampling of the latent space and the requested elastic properties to the corresponding design of the unit cell. The reconstruction part of CVAE is separated and is referred to as the "unit cell generative model" to generate RN unit cells given the target elastic properties. The elastic properties of the generated unit cells are further predicted by the unit cell elastic properties model for final filtering. (b) A demonstration of the ability of the unit cell generative model to generate new unit cells with given elastic properties which were not present in the initial library. Cross-sections are presented to more clearly visualize the generated unit cells with new elastic properties. (c) The deformation pattern of three specimens (*i.e.*, I, II, III) with new elastic properties not present in the original library. Moreover, a group of specimens (*i.e.*, IV) are presented to show the ability of the trained model to generate specimens with similar elastic properties.

**Figure 3.** A schematic illustration and elastic properties of the combinatorial designs as well as the network architecture of the size-agnostic model. (a) Each combinatorial design with a given dimension ($D_1 \times D_2$) is created by repeat filliping a specific RN unit cell $D_1$ times vertically (along direction 1) and $D_2$ times horizontally (along direction 2). The vectors containing $D_1$ and $D_2$ and the binary vectors representing the design of the unit cells are introduced to the size-agnostic model as input. The model then returns the predicted elastic properties of the combinatorial design as output. (b) The envelope of the elastic properties achieved by the combinatorial designs, according to direct FE simulations. (c) The evolution of the elastic properties as functions of $D_1$ and $D_2$ for a specific case study. (d) The deformation patterns of four combinatorial designs subjected to the following loading conditions: $\varepsilon_{11} = 5\%$ and $\varepsilon_{22} = 5\%$. Both the results of FE analysis and experimental results obtained using mechanical tests on the 3D printed specimens are presented.

**Figure 4.** The structure of Deep-DRAM as a size-agnostic inverse design framework. (a) In this framework, the deep generative model and the size-agnostic model are combined to generate combinatorial designs with desired elastic properties and dimensions. The best candidates among the generated combinatorial designs are then selected based on their MSE values. (b) The heat maps of the MSE values indicating the expected error values for generating combinatorial designs with predefined elastic properties and dimensions.

**Figure 5.** Multi-objective design where the minimization of peak von Mises stresses is considered as an additional design requirement. (a) Finding the optimized combinatorial designs based on the stress values of the elements of the lattice structure. (b) As representative cases, combinatorial designs of three groups (*i.e.*, $D_1 = D_2 = 4$, $D_1 = 10$ and $D_2 = 4$, and $D_1 = D_2 = 10$) were studied. For each group, the first 1000 generated combinatorial designs with MSE values below 0.1 were further analyzed using FE simulations to determine the stress distribution within their elements. The normalized peak von Mises stresses when the structure was subjected to a strain of $\varepsilon_{11} = 5\%$ are plotted against the same type of stress when the applied stain is $\varepsilon_{22} = 5\%$. (c) The deformation patterns and stress distributions of the elements of some selected combinatorial designs.

**Figure 1**

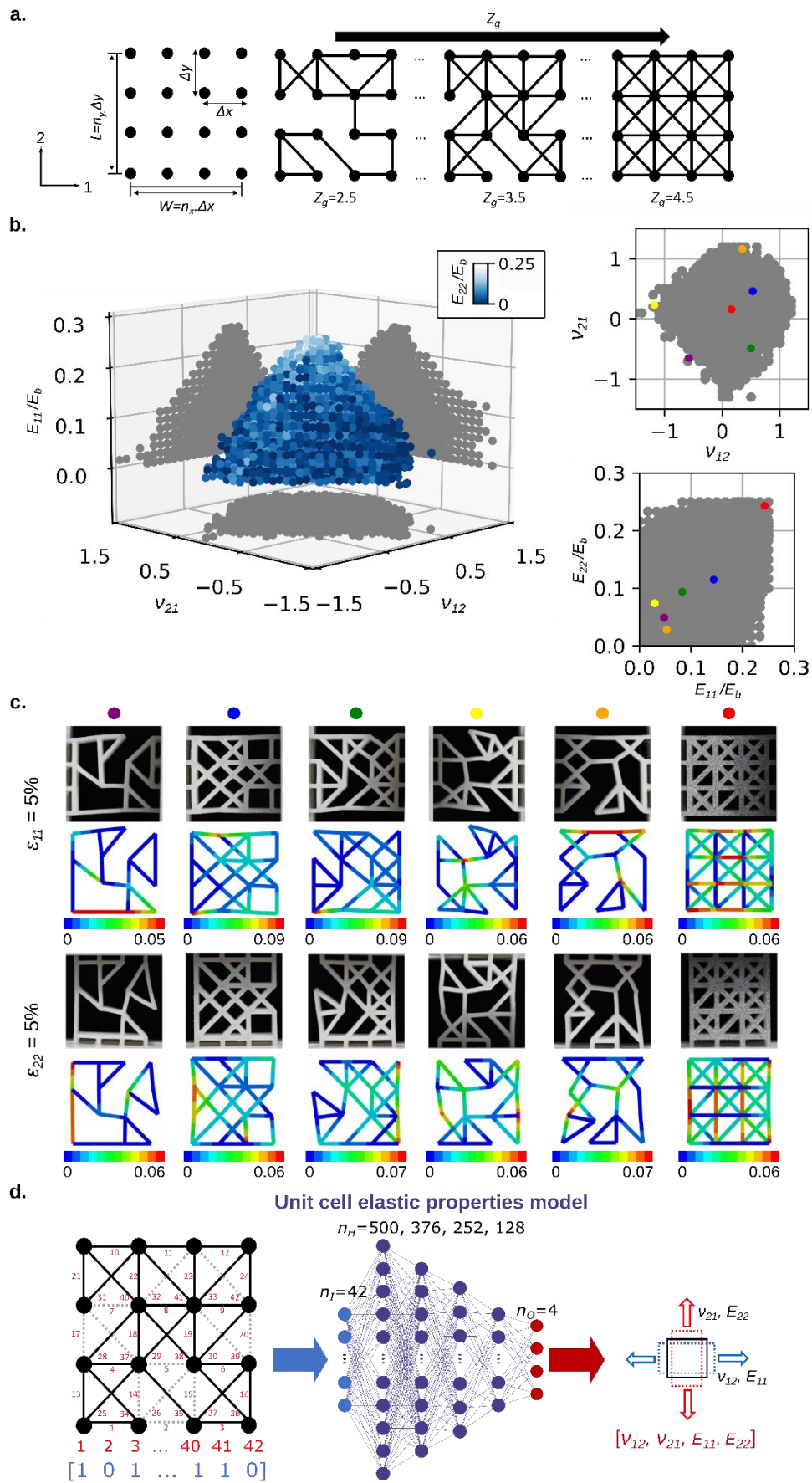

**Figure 2**

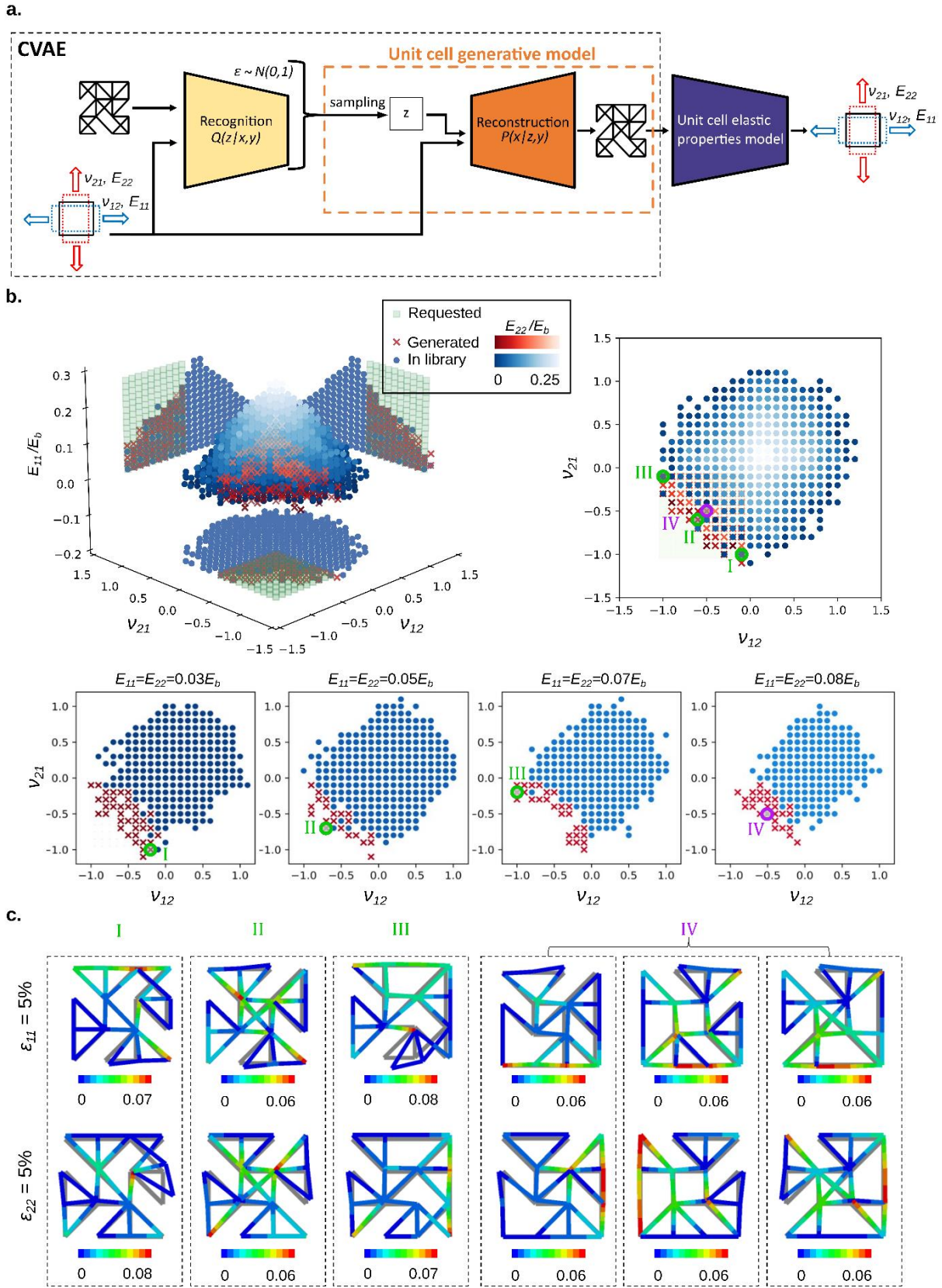

**Figure 3**

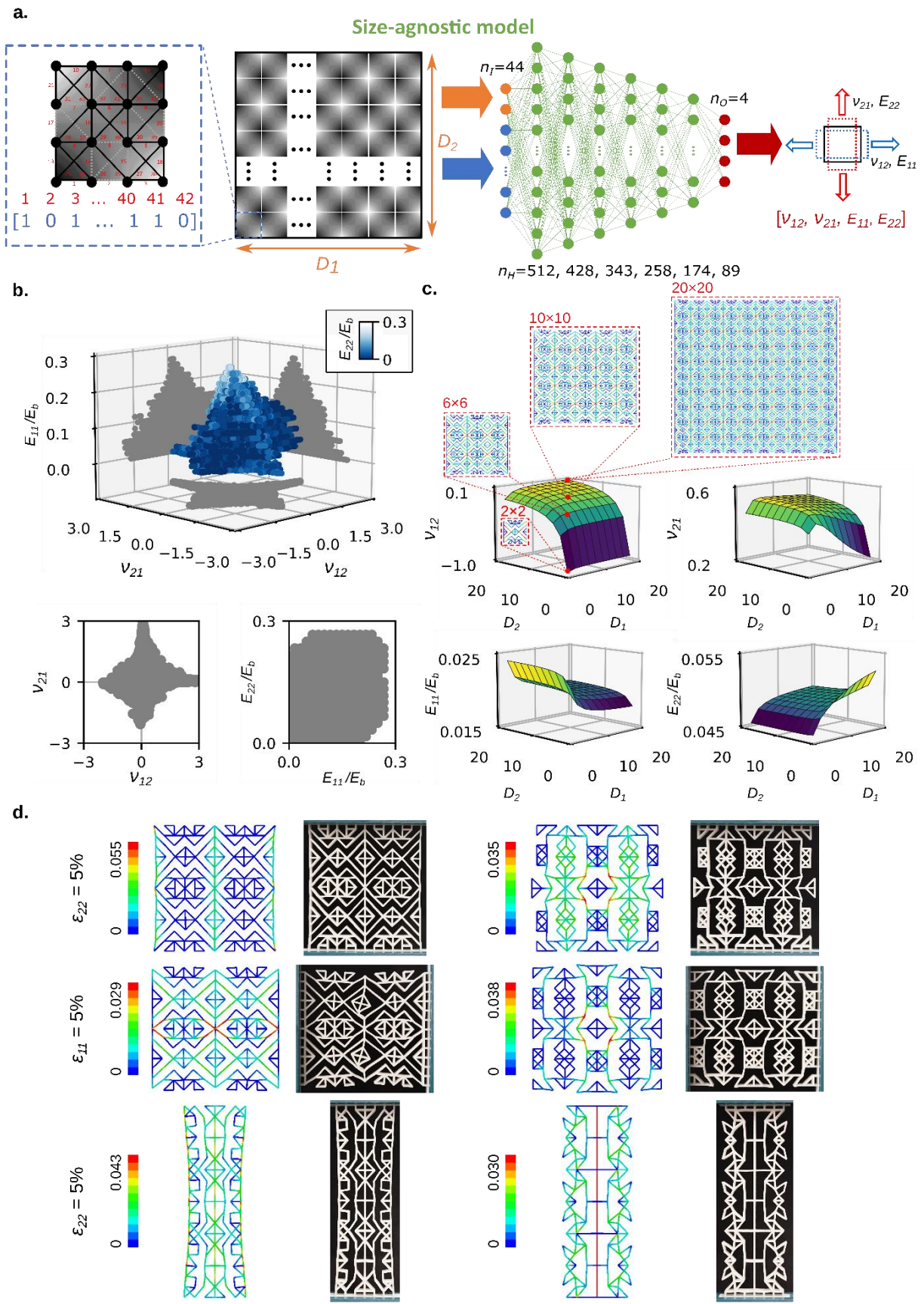

**Figure 4**

**a.**

**b.**

# Figure 5

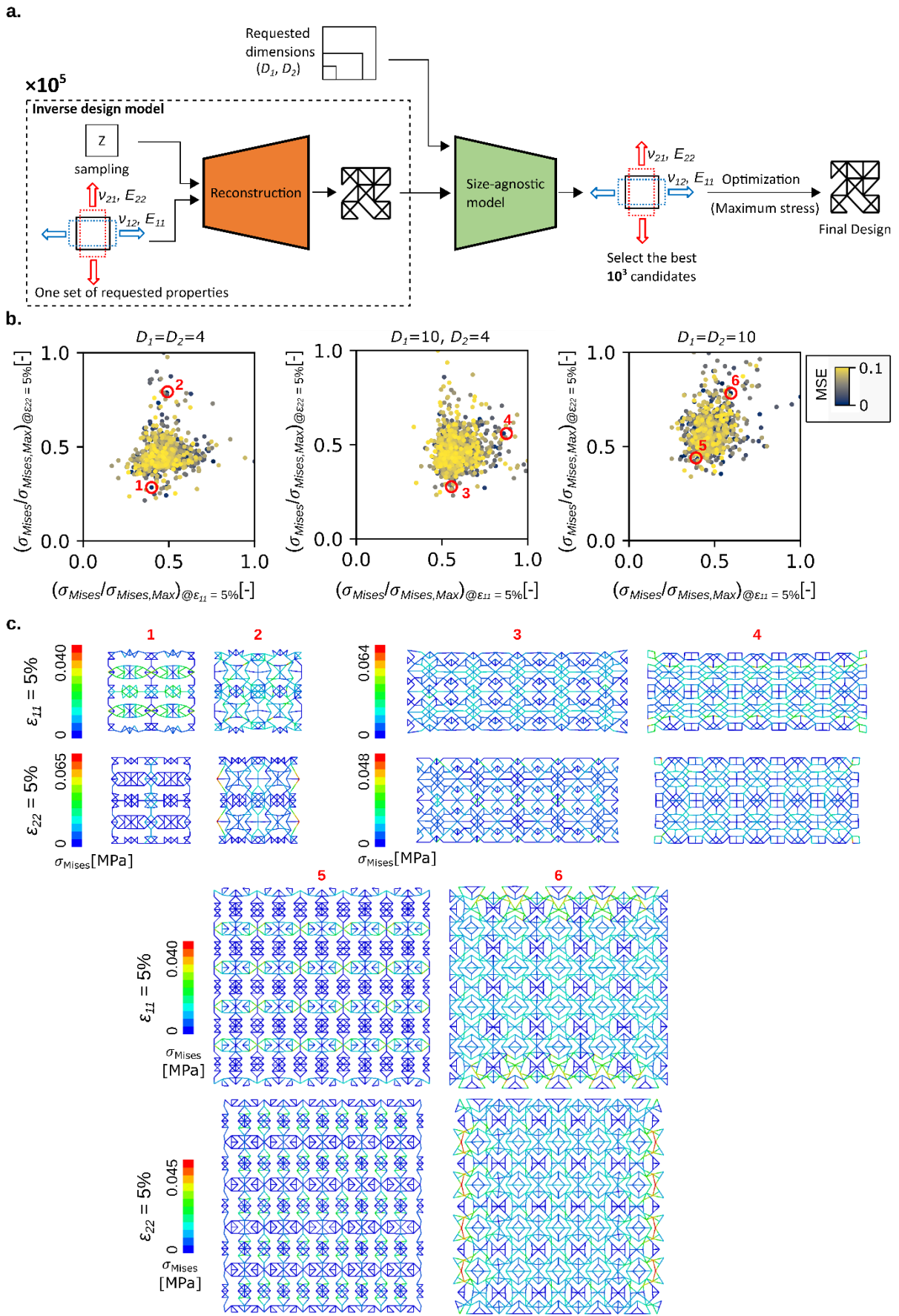

Supplementary document to

# Deep learning for size-agnostic inverse design of random network 3D-printed mechanical metamaterials

H. Pahlavani[a,2], K. Tsifoutis-Kazolis[a], P. Mody[b], J. Zhou[a], M. J. Mirzaali[a], A. A. Zadpoor[a]

[a] *Department of Biomechanical Engineering, Faculty of Mechanical, Maritime, and Materials Engineering, Delft University of Technology (TU Delft), Mekelweg 2, 2628 CD, Delft, The Netherlands*

[b] *Divsion of Image Processing (LKEB), Radiology, Leiden University Medical Center (LUMC), Albinusdreef 2, 2333 ZA, Leiden, The Netherlands*

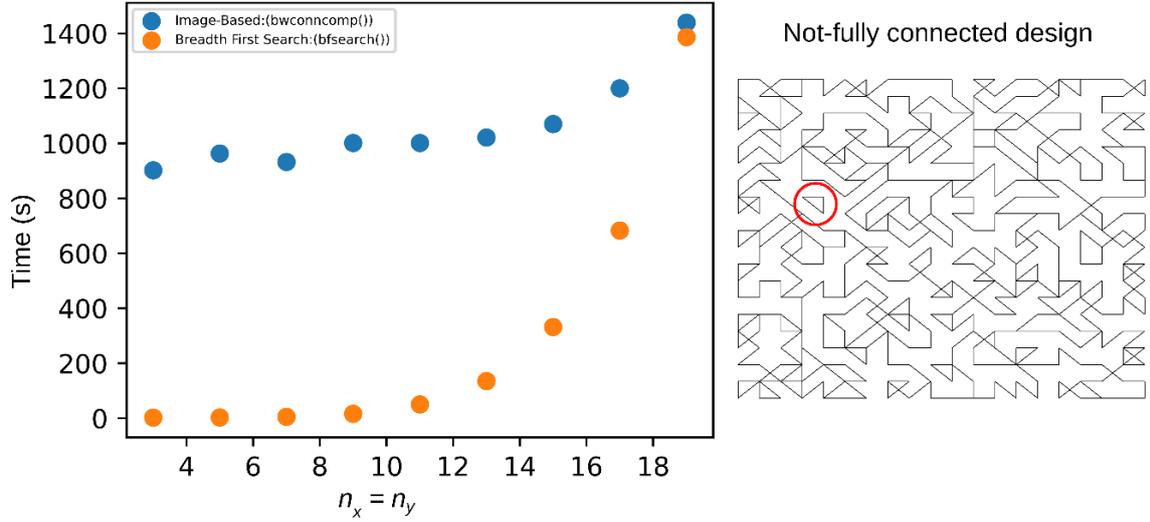

**Supplementary Figure 1.** Random network design filtering. This figure illustrates how the connectivity of RN unit cells are checked. A comparison between two different algorithms for the detection of non-connected nodes and their time efficiency. The analysis is performed for different node sizes ($n_x = n_y$).

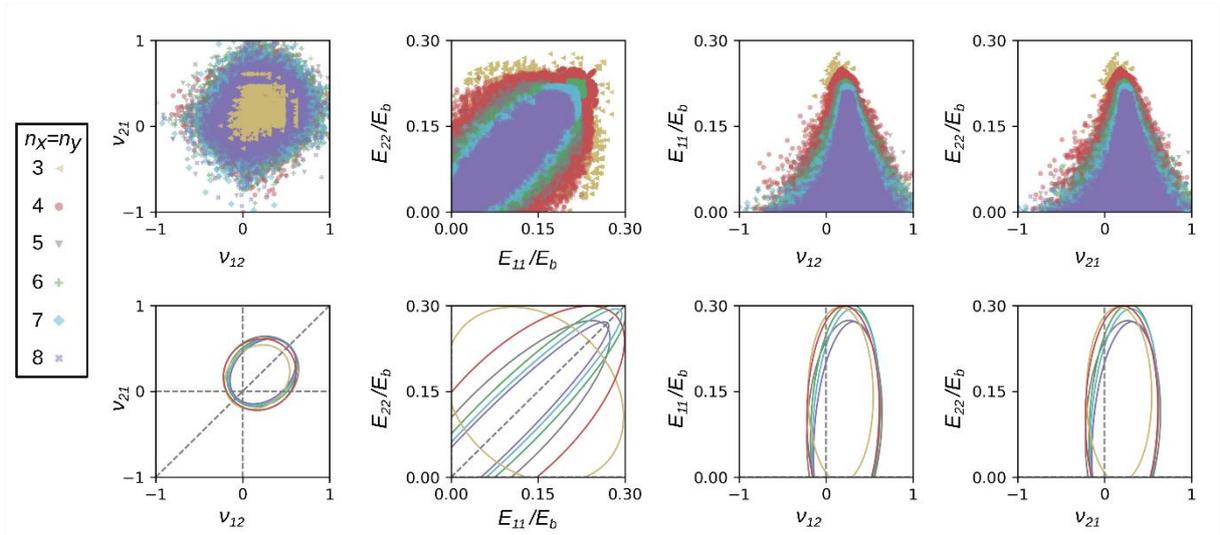

**Supplementary Figure 2.** The elastic properties (top row) and the confidence ellipses (bottom row) for RN unit cells with different node numbers (i.e., $n_x = n_y = 3, 4, 5, 6, 7,$ and $8$). $E_b$ is the elastic modulus of the bulk material

**Supplementary Table 1.** The principal radii ($r_1$ and $r_2$) as well as the areas ($A$) of the confidence ellipses presented in Supplementary Figure 2.

| $n_x = n_y$ | $\nu_{12}, \nu_{21}[-]$ | | | $E_{11}/E_b, E_{22}/E_b[-]$ | | | $\nu_{12}, E_{11}/E_b[-]$ | | | $\nu_{21}, E_{22}/E_b[-]$ | | |
|---|---|---|---|---|---|---|---|---|---|---|---|---|
| | $r_1$ | $r_2$ | $A$ | $r_1$ | $r_2$ | $A$ | $r_1$ | $r_2$ | $A$ | $r_1$ | $r_2$ | $A$ |
| 3 | 0.43 | 0.41 | 0.55 | 0.17 | 0.15 | 0.08 | 0.38 | 0.16 | 0.19 | 0.38 | 0.16 | 0.19 |
| 4 | 0.46 | 0.41 | 0.59 | 0.24 | 0.11 | 0.08 | 0.42 | 0.19 | 0.24 | 0.42 | 0.19 | 0.24 |
| 5 | 0.45 | 0.37 | 0.52 | 0.25 | 0.08 | 0.06 | 0.43 | 0.18 | 0.25 | 0.43 | 0.18 | 0.25 |
| 6 | 0.43 | 0.34 | 0.46 | 0.27 | 0.06 | 0.05 | 0.41 | 0.19 | 0.25 | 0.41 | 0.19 | 0.25 |
| 7 | 0.43 | 0.32 | 0.43 | 0.29 | 0.05 | 0.04 | 0.39 | 0.2 | 0.24 | 0.39 | 0.2 | 0.25 |
| 8 | 0.41 | 0.35 | 0.45 | 0.27 | 0.04 | 0.03 | 0.38 | 0.18 | 0.22 | 0.38 | 0.18 | 0.22 |

**Supplementary Table 2.** The probability of finding different ranges of the Poisson's ratio (%) for datasets with different $n_x$ and $n_y$ values.

| $n_x = n_y$ | $\nu_{12}, \nu_{21} < -0.5$ | $-0.5 \leq \nu_{12}, \nu_{21} < 0$ | $0 \leq \nu_{12}, \nu_{21} < 0.5$ | $(\nu_{12}, \nu_{21} \geq 0.5)$ |
|---|---|---|---|---|
| 3 | 0.00 | 6.75 | 92.75 | 0.50 |
| 4 | 0.05 | 6.63 | 90.47 | 2.84 |
| 5 | 0.05 | 6.01 | 91.43 | 2.50 |
| 6 | 0.04 | 6.00 | 91.47 | 2.49 |
| 7 | 0.03 | 5.43 | 92.51 | 2.02 |
| 8 | 0.03 | 5.17 | 93.02 | 1.78 |

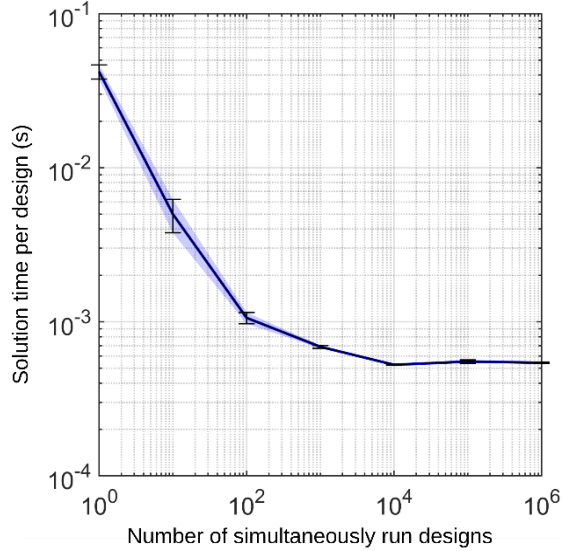

**Supplementary Figure 3.** Time comparison for the direct FE simulations aimed at determining the elastic properties of the RN unit cells.

**Supplementary Table 3.** The estimated time (considering $5.4 \times 10^{-4}$ s for each RN unit cell) to generate all possible RN unit cells with $n_x = n_y = 4$. The estimated number of RN unit cells is calculated as $C_{El_{placed}}^{El_{all}} = \frac{El_{all}!}{(El_{all}-El_{placed})! \, El_{placed}!}$, where, $El_{all} = n_x(n_y - 1) + n_y(n_x - 1) + 2(n_x - 1)(n_y - 1)$, $El_{placed} = \frac{1}{2} Z_g \times n_x \times n_y$.

| $Z_g$ | The estimated number of RN unit cells | Estimated Time [s] |
|---|---|---|
| 2.5 | $C_{20}^{42} \approx 5.14 \times 10^{11}$ | 277,447,468 |
| 3 | $C_{24}^{42} \approx 3.54 \times 10^{11}$ | 190,996,445 |
| 3.5 | $C_{28}^{42} \approx 5.29 \times 10^{10}$ | 28,544,524 |
| 4 | $C_{32}^{42} \approx 1.47 \times 10^{09}$ | 794,579 |
| 4.5 | $C_{36}^{42} \approx 5.25 \times 10^{06}$ | 2,832 |
| Total | $9.22 \times 10^{11}$ | 497,785,848 |

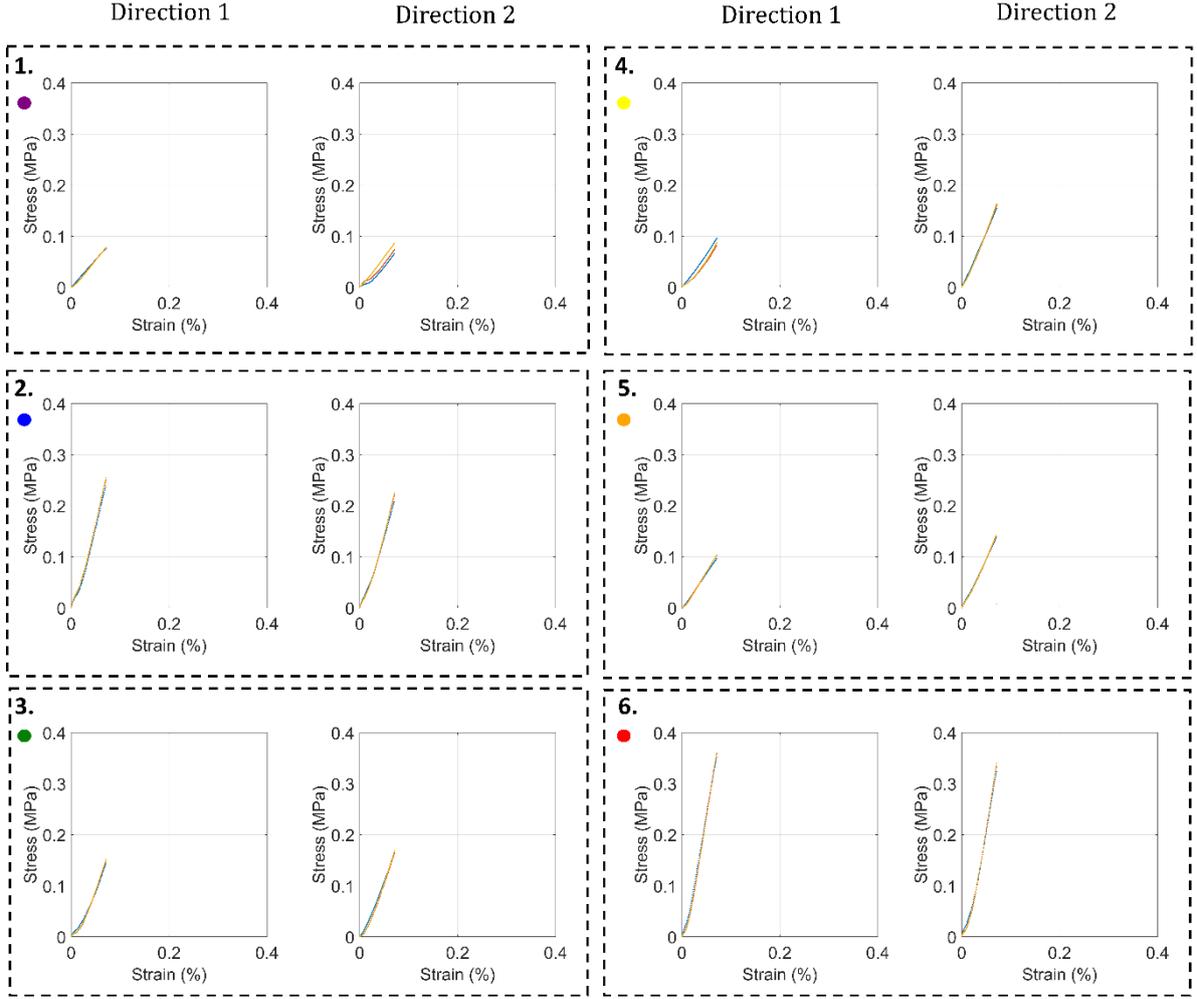

**Supplementary Figure 4.** The stress-strain curves for the experimental results of the RN unit cells. For the color code of each specimen see Figures 1b and 1c.

**Supplementary Table 4.** The numerical simulations and experimental results for the elastic properties of the RN unit cells. The numerically calculated elastic moduli are normalized to the elastic modulus of the bulk material assumed in the FE analysis ($E_{b,FE} = 0.6$ MPa) while the experimentally obtained elastic moduli are normalized with respect to the elastic modulus of the bulk material from which the samples were printed ($E_{b,EXP} = 25$ MPa).

| Sample | Numerical Simulations | | | | Experimental Results | | | |
|---|---|---|---|---|---|---|---|---|
| | $v_{12}[-]$ | $v_{21}[-]$ | $E_{11}/E_{b,FE}[-]$ | $E_{22}/E_{b,FE}[-]$ | $v_{12}[-]$ | $v_{21}[-]$ | $E_{11}/E_{b,EXP}[-]$ | $E_{22}/E_{b,EXP}[-]$ |
| 1 | −0.58 | −0.65 | 0.048 | 0.049 | −0.39 | −0.48 | 0.048 | 0.045 |
| 2 | 0.53 | 0.46 | 0.144 | 0.115 | 0.59 | 0.41 | 0.152 | 0.133 |
| 3 | 0.50 | −0.49 | 0.083 | 0.094 | 0.20 | −0.41 | 0.091 | 0.104 |
| 4 | −1.18 | 0.23 | 0.030 | 0.074 | −0.88 | 0.54 | 0.053 | 0.097 |
| 5 | 0.35 | 1.16 | 0.053 | 0.028 | 0.39 | 0.86 | 0.084 | 0.063 |
| 6 | 0.16 | 0.16 | 0.243 | 0.243 | 0.31 | 0.27 | 0.229 | 0.209 |

## METHODS

### 1. Finite Element modeling

The element stiffness matrix was transferred to the global coordinate ($K^e$) and was calculated as[1,2]:

$$K^e = Q^T \bar{K}^e Q, \tag{1}$$

$$\bar{K}^e = \frac{E_b}{(1+\mu)} \begin{bmatrix} A(1+\mu)/L_e & 0 & 0 & -A(1+\mu)/L_e & 0 & 0 \\ 0 & 12I/L_e^3 & 6I/L_e^2 & 0 & -12I/L_e^3 & 6I/L_e^2 \\ 0 & 6I/L_e^2 & 4I(1+\mu/4)/L_e & 0 & -6I/L_e^2 & 2I(1-\mu/2)/L_e \\ -A(1+\mu)/L_e & 0 & 0 & A(1+\mu)/L_e & 0 & 0 \\ 0 & -12I/L_e^3 & -6I/L_e^2 & 0 & 12I/L_e^3 & -6I/L_e^2 \\ 0 & 6I/L_e^2 & 2I(1-\mu/2)/L_e & 0 & -6I/L_e^2 & 4I(1+\mu/4)/L_e \end{bmatrix}, \tag{2}$$

$$\mu = \frac{12 E_b I}{L_e^2 G_b A K_s}, \tag{3}$$

$$Q = \begin{bmatrix} n_{x\bar{x}} & n_{y\bar{x}} & 0 & 0 & 0 & 0 \\ n_{x\bar{y}} & n_{y\bar{y}} & 0 & 0 & 0 & 0 \\ 0 & 0 & 1 & 0 & 0 & 0 \\ 0 & 0 & 0 & n_{x\bar{x}} & n_{y\bar{x}} & 0 \\ 0 & 0 & 0 & n_{x\bar{y}} & n_{xy} & 0 \\ 0 & 0 & 0 & 0 & 0 & 1 \end{bmatrix}, \tag{4}$$

where $\bar{K}^e$ is the local element stiffness matrix, and $E_b$, $A$, $I$, and $L_e$ are the elastic modulus of the bulk material, the cross-section area ($A = Tt$) of the element, the moment of inertia ($I = Tt^3/12$) of the element, and the length of the element, respectively. $\mu$ is a dimensionless coefficient that characterizes the importance of shear-related parameters including $G_b$ (shear modulus of the bulk material) and $K_s$ (shear correction factor = 0.85). $Q$ is the transformation matrix and contains the direction cosines:

$$n_{x\bar{x}} = n_{y\bar{y}} = \frac{x_2 - x_1}{L_e}, n_{y\bar{x}} = -n_{x\bar{y}} = \frac{y_2 - y_1}{L_e} \tag{5}$$

where $x_1, y_1, x_2$, and $y_2$ are the element nodal coordinates. The stiffness matrix was calculated for all the elements and was assembled into a global stiffness matrix ($K$).

The predefined boundary conditions for our displacement control ($f = 0$) analysis are applied as the following boundary condition matrix:

$$bc = \begin{bmatrix} DOF_1 & (u_{11})_{node\_1} \\ DOF_2 & (u_{22})_{node\_1} \\ DOF_3 & (u_{33})_{node\_1} \\ \vdots & \vdots \\ DOF_{3 \times n} & (u_{33})_{node\_n} \end{bmatrix} \tag{6}$$

where the first column shows the number of DOF and the second column shows the corresponding predefined displacements. Considering $n$ as the total number of nodes and three DOF for each node, we had $3 \times n$ DOF in total. Finally, we used solveq function ($[a, r] = solveq(K, f, bc)$) of CALFEM finite element toolbox[2] to calculate reaction forces ($r$) and displacements ($a$) in all DOF.

## 2. Hyperparameter tuning pipeline

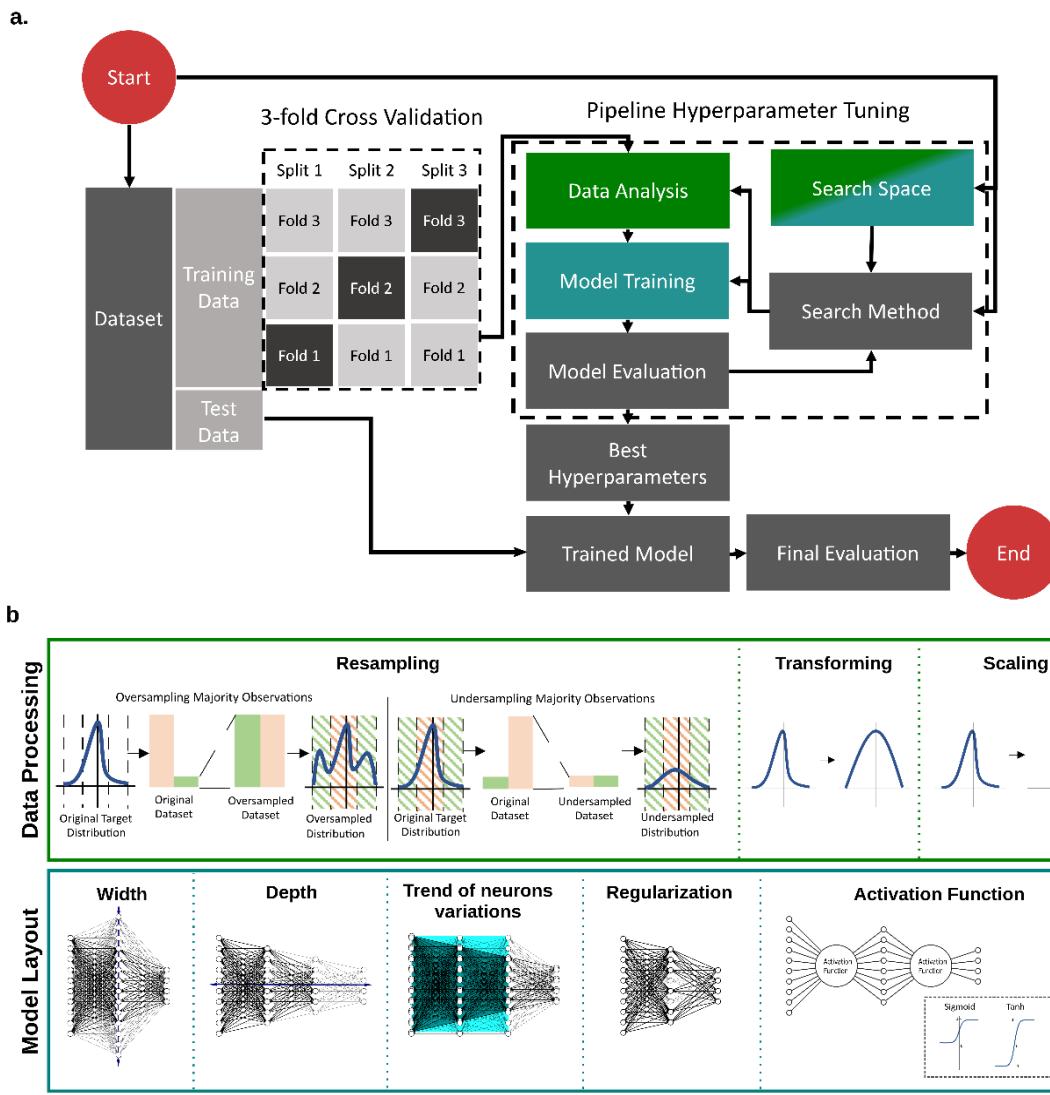

**Supplementary Figure 5.** The training pipeline. a) The suggested methodology for the training of deep learning models while avoiding the pitfalls of common practice and introducing an extra step of optimization in the search of best parameters, b) The data analysis and model hyperparameter options for the search space of the hyperparameter tuning procedure.

### 2.1. Data processing

Various data processing methods can aid in the successful training of a deep learning model. The features have binary values and no further analysis is, therefore, required. The distributions of the targets were acquired. It is noticeable that the distribution of the Poisson's ratio for both directions have heavy tails, explained by the rarity of the specimens with extreme negative and positive values of the Poisson's ratios. Furthermore, the Poisson's ratios and elastic moduli have different ranges of values (*i.e.*, $v_{12}$, $v_{21} = [-1.5, 1.5]$ and $E_{11}/E_b, E_{22}/E_b = [0, 0.25]$).

Model training on highly skewed data tends to be misleading and leads to poor model performance[3]. These observations within the tail of the distributions refer to rare but useful data samples, such as double-auxeticity or strong auxeticity. In this case, the dataset is imbalanced with some data of interest underrepresented. To balance out the dataset based on the targets, a data resampling process is proposed. Prior to training a deep learning model, the input and output data can be prepared by utilizing transformation/rescaling techniques[3]. Models
30

commonly work more efficiently when the inputs have a normal distribution, and it is also shown that the transformation of the outputs could help in improving the performance of the model[3]. An input with a variance that is orders of magnitude greater than others may dominate the objective function and prevent the estimator from learning from other inputs. Output transformation may also aid in the training of the model. Some form of output scaling is also suggested to make the model converge more easily, as unscaled output variables can cause exploding gradients when dealing with regression problems[3]. Taking into consideration the distribution and the range of the targets, the training dataset could be processed accordingly by including the option of target-based data resampling and transformation/rescaling options (either as output for the forward models or as input for the inverse model) in the search space of the hyperparameter tuning pipeline.

**Data resampling:** Addressing classification issues for imbalanced data is well-documented. However, regression training on imbalanced data is not as widely studied. Using the same techniques as in other cases is often problematic. Pre-processing approaches are primarily concerned with oversampling the minority samples, undersampling the majority ones, or a combination of both before training the regression model[4]. Oversampling depends on duplicating some data that do not have enough representation or fabricating similar data. The simplest technique for oversampling is to produce new samples by randomly sampling and replacing existing samples[4]. Undersampling cuts down the total number of the available data points, selecting a roughly similar size of each type of data that needs to be represented[4]. For oversampling, we used the Synthetic Minority Over-Sampling Technique for Regression with Gaussian Noise (SMOGN)[5], which is beneficial for regression problems where the prediction of rare values is of great importance.

By undersampling the dataset, the model can be trained faster while having equivalent information to work with. Undersampling for classification is a much easier work due to the explicit segmentation of classes that is specified from the beginning. However, in the case of a continuous variable, the areas of resampling are more difficult to specify. The binning of the dataset was performed by manually selecting the number of "classes" to obtain datasets with a similar number of data points. As a preliminary inspection of bin-based undersampling, three bins were created for the following ranges of the Poisson's ratio: $-1.5 < v < 0$, $0 < v < 0.3$, and $0.3 < v < 1.5$. The second class appeared to be the most common while the other two were underrepresented. The results revealed that the initial dataset was greatly downsized. To better classify the dataset, samples with relatively similar elastic properties were grouped together. The number of samples per each group was then kept constant. The values of $E$ and $v$ were rounded to the first and second decimals respectively, as smaller variations of the elastic properties can be neglected. In that way, multiple tiles can be grouped to have similar elastic properties. The frequency of each unique combination of the four elastic properties was then calculated. It is clear that there are more duplicates for specific elastic properties. Finally, we kept five unit cells for each pair of the elastic properties when available. Consequently, a dataset of six million-unit cells reduced to eighty thousand unit cells. Both oversampling and undersampling were included in the search space considered for hyperparameter tuning.

**Data transformation/scaling:** Scaling features to be confined in the range between specific minimum and maximum values is an alternative standardization method. In the case of multitask regression, normalizing all the targets aids in balancing the training loss from various individual tasks. The rescaling through MinMaxScaler and the quantile transformation were included in the search space of the hyperparameter tuning pipeline applied before training the model. MinMaxScaler transforms features by scaling each feature such that it is in the range of [0,1]. Quantile transformer is a class that transforms features and targets to fit a uniform or normal distribution, given enough training examples.

## 2.2. Model hyperparameters

Training of a deep learning model requires the modification of several hyperparameters. Here, we tuned the architecture of the model (including the number of hidden layers, the number of hidden neurons per each layer, and the trend of the changes of the neurons of different hidden layers), the activation function, the optimizer, and the regularization terms as hyperparameters. Then, we created the models with the use of Keras TensorFlow API for Sequential Models. The search methods of cross-validation grid search from scikit-learn[3] were then used to systematically iterate among a library of predefined values for the hyperparameters.

**Capacity/architecture of the model:** Each model was created from scratch. The architecture of a neural network model is modified 1) by the width, determined by the number of neurons in each layer and 2) the depth which is determined by the number of layers, and 3) the trend of the changes of the neurons of different hidden layers. The number of neurons in the input layer equals the number of features in the data and the number of neurons in the output layer for supervised learning equals the number of targets/labels. As there is no predefined formula that guarantees the high accuracy of a model without overfitting, we systematically study different combinations of hyperparameters. The parameterized sequential model was built based on a function that takes the following as its inputs: the number of hidden layers, the input and output size, the width of the first hidden layer, and the trend of the variation of the number of hidden neurons per layer (whether the number of neurons remains the same throughout the hidden layers or is it gradually decreasing). The width of the other hidden layers is declining by a specific step which is defined as the difference between the number of neurons of the first hidden and output layer, divided by the number of hidden layers. For the CVAE, a final parameter that affects the capacity of the models is the dimension of the latent space. The bottlenecking was expected to affect the ability of the reconstruction/inverse model. Therefore, various values [2, 8, and 16] were considered for the dimension of the latent space.

**Regularization**: When there is no way to expand the data, overfitting can be combated by regularization, which is a way to constrain the amount and kind of information that the model can hold, forcing the model to focus on the most essential parameters to achieve generalization. The most prevalent regularization approaches are weight regularization and dropout. Batch normalization can also operate as a regulator, decreasing and, in certain cases, eliminating the requirement for dropout. The option of batch normalization inclusion after each hidden layer was, thus, included in the hyperparameter optimization pipeline.

**Optimizers:** Optimizers are algorithms that are used to minimize losses by adjusting the characteristics of the neural network, such as weights and learning rate. An optimization algorithm called gradient descent is frequently used to train deep learning models. In current deep learning algorithms, three forms of gradient descent learning algorithms are used: batch gradient descent, stochastic gradient descent, and small batch gradient descent. The simplest yet widely used optimization approach is batch gradient descent. It is extensively utilized in linear regression and classification techniques. However, for big datasets, batch gradient descent performs redundant computations since it recomputes gradients for comparable cases before each parameter change. By completing one update at a time, stochastic gradient descent (SGD) eliminates this redundancy. As a result, it is significantly faster. Many methods have been proposed for optimizing SGD, among which we selected Adam and RMSprop optimizers to be evaluated through the hyperparameter tuning pipeline.

**Activation function:** Activation functions are widely utilized because of a few desired properties: non-linearity, range, and differentiability. For a universal function approximator, nonlinear activations are desirable. Moreover, the function should be continuously differentiable for gradient-based optimization approaches to be possible. The model's stability and efficiency are influenced by whether the function's range is finite or not. Sigmoid, tanh, and ReLU are examples of popular activation functions that fit the selection criteria and were included as options in the search space for creating the sequential neural network models. The activation functions of hidden layers and the output layer were defined separately. If the range of the output activation function is shorter than the range of the dataset values, a value cut-off can be detected. For the output layer of the reconstruction model of the CVAE, the activation function was pre-set to sigmoid, to output a binary probabilistic representation of the vector that describes the designs, which would allow the reconstruction part of the overall CVAE loss to be approximated with the binary cross entropy.

**Batch size:** The batch size is an important hyperparameter, as there is a relation between the batch size on the one hand and the speed and stability of the learning process on the other. The recommended batch size varies depending on whether the learning method uses batch, stochastic, or minibatch gradient descent. Stochastic gradient descent algorithms that are selected work in a small-batch mode, sampling a subset of the training data, typically 32 to 512 data points. In fact, it has been shown that employing a bigger batch dramatically lowers the model's quality as measured by its capacity to generalize[6]. Thus, the batch size was set at 32.

**Supplementary Table 5.** The search space of the hyperparameter tuning pipeline as applied to the unit cell elastic properties model.

| Step | Process | Search Space | Selected |
|---|---|---|---|
| Data Analysis | Resampling | Undersampling, Oversampling | Undersampling |
| | Rescaling/transforming | None, MinMaxScaler, Quantiletransformer | MinMaxScaler |
| Model Tuning | Hidden layers | $2 - 5$ | 4 |
| | Neurons $1^{st}$ hidden layer | $20, 100, 500$ | 500 |
| | Trend of neurons variations | Rectangle, Triangle | Triangle |
| | Activation function | None, Relu, Tanh | Relu |
| | Last layer activation function | None, Relu, Tanh | Relu |
| | Optimizer | Adam, RMSprop | Adam |
| | Step size | 0.0001 | |
| | Loss function | MSE | |
| | Batch normalization | Yes, No | No |
| | Targets | 4 | |
| Fitting | Batch size | 32 (Default) | |
| | Epochs | 200 | |
| Cross-validation | K-folds | 3 | |
| | Data split | $60 - 30 - 10$ | |
| | Scoring | $R^2$ | |

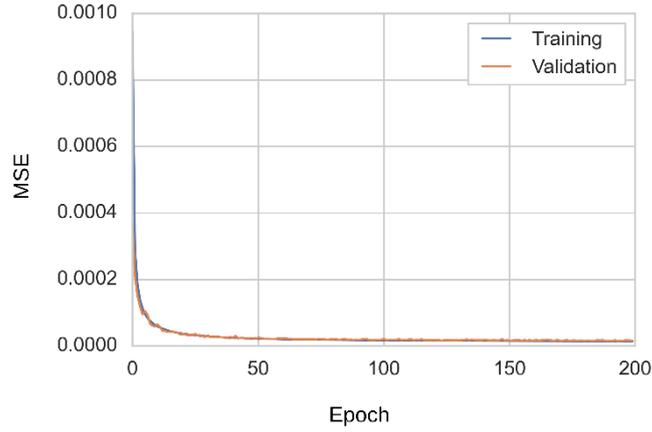

**Supplementary Figure 6.** The MSE graph for the unit cell elastic properties model.

**Supplementary Table 6.** The regression metrics. The evaluation of the unit cell elastic properties model using the test dataset. In this table, $R^2$ is the coefficient of determination ($R^2 = 1 - \frac{\sum_{i=1}^{n}(y_i - \hat{y}_i)^2}{\sum_{i=1}^{n}(y_i - \bar{y})^2}$), $MSE$ is the mean squared error ($MSE = \frac{1}{n}\sum_{i=1}^{n}(y_i - \hat{y}_i)^2$), $MAE$ is the mean absolute error ($MAE = \frac{1}{n}\sum_{i=1}^{n}|y_i - \hat{y}_i|$), and $RMSE$ is the root mean squared error ($RMSE = \sqrt{MSE}$), where $n$ is the size of the dataset, $y_i$ is the $i^{th}$ real target, $\hat{y}_i$ is the corresponding predicted value, and $\bar{y}$ is the mean value of $y$ ($\bar{y} = \frac{1}{n}\sum_{i=1}^{n} y_i$).

| Index | $v_{12}$ | $v_{21}$ | $E_{11}$ | $E_{22}$ | Overall |
|---|---|---|---|---|---|
| $R^2$ | 0.9934 | 0.9945 | 0.9997 | 0.9997 | 0.9968 |
| MSE | 0.0001 | 0.0001 | 0.0000 | 0.0000 | 0.0000 |
| MAE | 0.0071 | 0.0065 | 0.0004 | 0.0004 | 0.0036 |
| RMSE | 0.0114 | 0.0104 | 0.0005 | 0.0005 | 0.0057 |

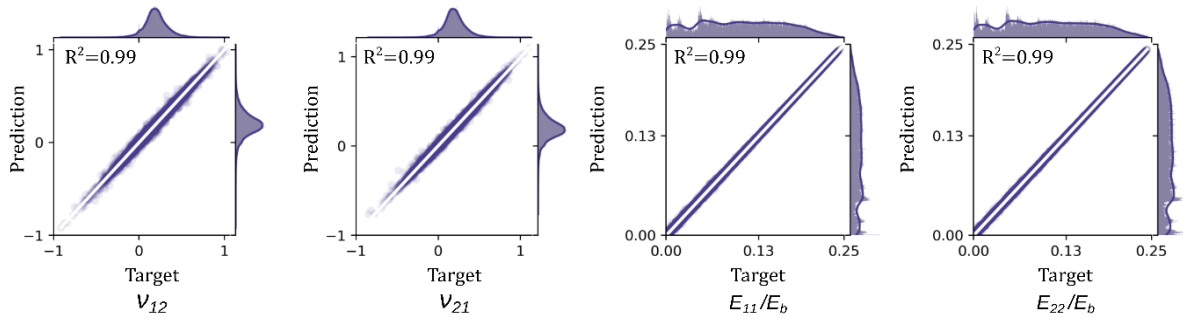

**Supplementary Figure 7.** The evaluation of the unit cell elastic properties model. The prediction *vs.* target values and the coefficients of determination for the test datasets.

**Supplementary Table 7.** The search space for the training of the CVAE model and the selected parameters.

| Step | Parameter | Search Space | Selected |
|---|---|---|---|
| Recognition Model | Inputs | Structures + elastic properties | |
| | Neurons $1^{st}$ hidden layer | 128, 512 | 512 |
| | Hidden layers | $1 - 4$ | 2 |
| | Trend of neurons variations | Rectangle, Triangle | Triangle |
| | Activation function | Relu, Sigmoid | Relu |
| | Last layer activation function | Relu, Sigmoid | Relu |
| | Outputs | Latent dimensions | |
| Latent Space Dimension | Size | 2, 8, 16 | 8 |
| Reconstruction Model | Inputs | Latent Dimensions + elastic properties | |
| | Neurons $1^{st}$ hidden layer | 128, 512 | 512 |
| | Hidden layers | $1 - 4$ | 2 |
| | Trend of neurons variations | Rectangle, Triangle | Triangle |
| | Activation function | Relu, Sigmoid | Relu |
| | Last layer activation function | Sigmoid | |
| | Outputs | Structure | |
| | Optimizer | Adam | |
| | Step size | 0.0001 | |
| | Loss function | $\mathcal{L}_{all} = \mathcal{L}_{CVAE} + \mathcal{L}_{MSE}$ | |
| | Batch normalization | No | |
| Fitting | Batch size | 32 (Default) | |
| | Epochs | 2000 | |
| Cross-validation | K-folds | 3 | |
| | Data split | $60 - 30 - 10$ | |
| | Scoring | $\mathcal{L}_{all} = \mathcal{L}_{CVAE} + \mathcal{L}_{MSE}$ | |

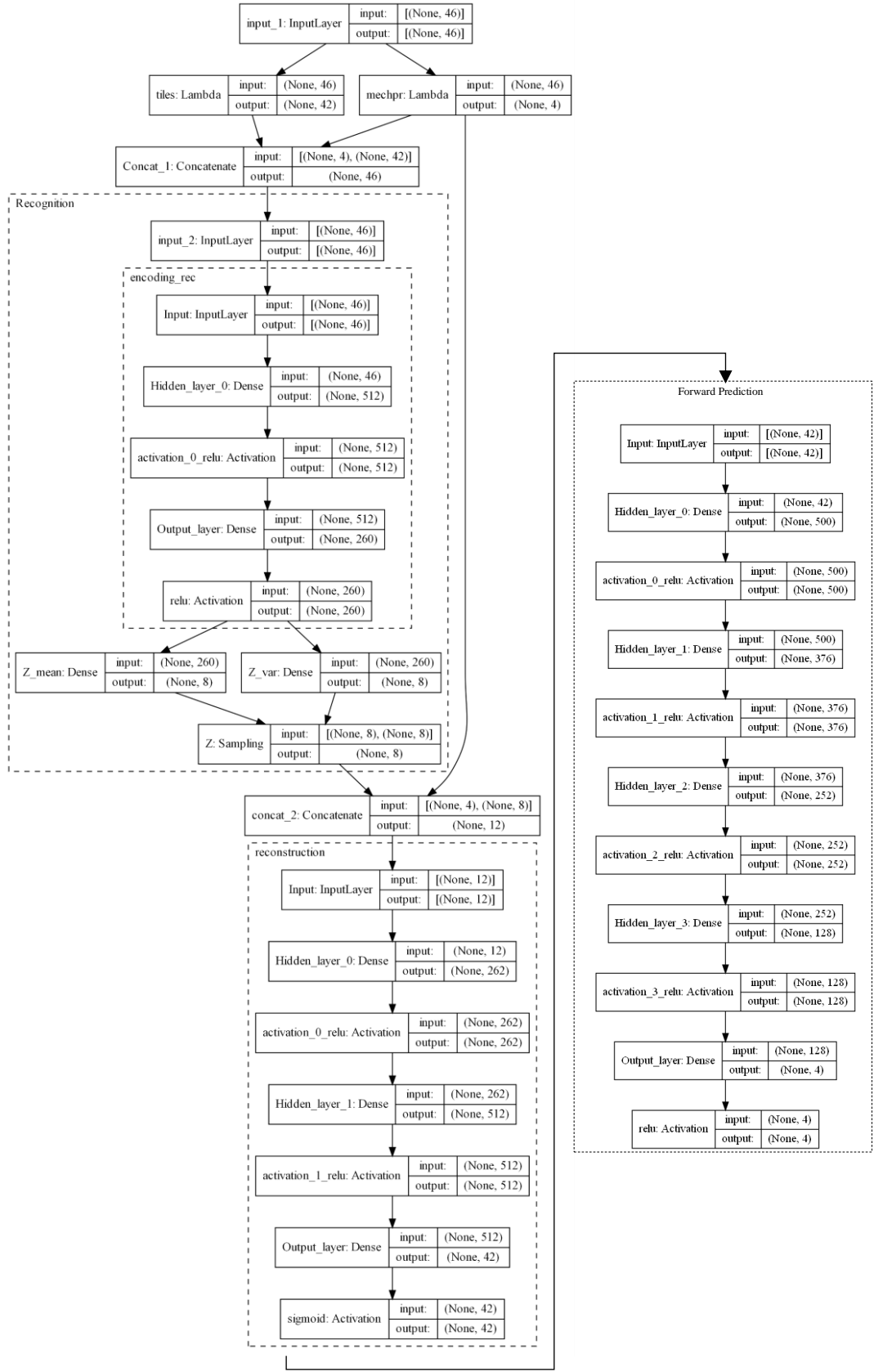

**Supplementary Figure 8.** The network architecture and selected hyperparameters of the CVAE model and forward prediction unit cell elastic properties model. The reconstruction model of the CVAE is then used to generate unit cells with given elastic properties.

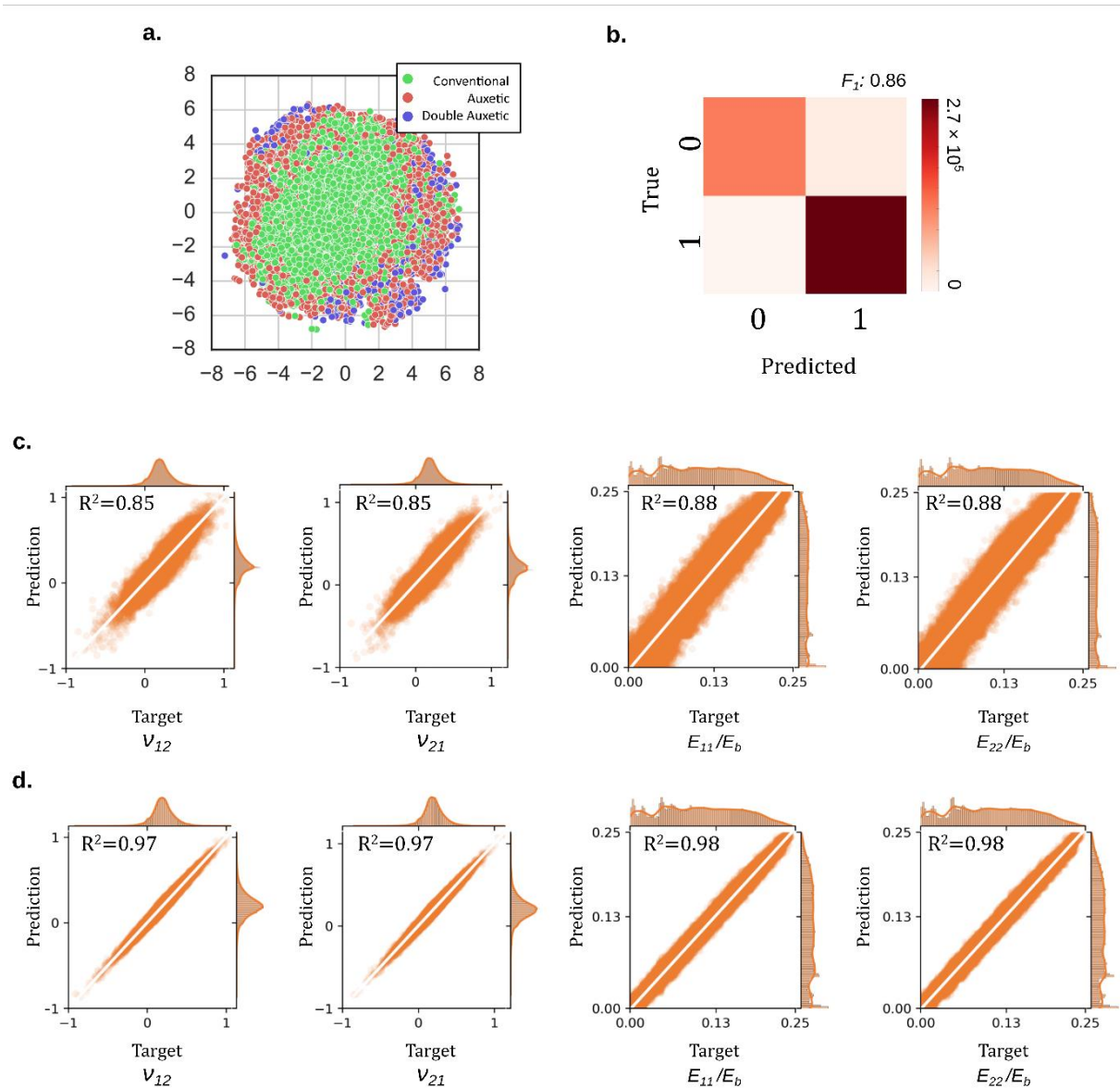

**Supplementary Figure 9.** The evaluation of the trained CVAE and the unit cell generative model. a) The *t*-SNE plot of the latent space created by the CVAE, displaying the placement of conventional, auxetic, and double auxetic metamaterials. b) The confusion matrix and $F1$ score for the reconstruction of the unit cell through the reconstruction model of the CVAE (*i.e.*, the unit cell generative model). c) The prediction error plot and $R^2$ for the predicted elastic properties of the designs generated by the unit cell generative model. d) The prediction error plot and $R^2$ for the predicted elastic properties of the best unit cell candidates from the unit cell generative model.

**Supplementary Table 8.** The regression metrics corresponding to the unit cell generative model for the test dataset, as displayed in Supplementary Figure 9c.

| Index | $v_{12}$ | $v_{21}$ | $E_{11}$ | $E_{22}$ | Overall |
|---|---|---|---|---|---|
| $R^2$ | 0.8460 | 0.8489 | 0.8878 | 0.8772 | 0.8650 |
| MSE | 0.0170 | 0.0168 | 0.0001 | 0.0001 | 0.0085 |
| MAE | 0.0935 | 0.0932 | 0.0086 | 0.0090 | 0.0511 |
| RMSE | 0.1305 | 0.1297 | 0.0011 | 0.0012 | 0.0710 |

**Supplementary Table 9.** Test regression metrics for the unit cell generative model as applied to the test dataset, displayed in Supplementary Figure 9d.

| Index | $v_{12}$ | $v_{21}$ | $E_{11}$ | $E_{22}$ | Overall |
|---|---|---|---|---|---|
| $R^2$ | 0.9672 | 0.9665 | 0.9870 | 0.9868 | 0.9769 |
| MSE | 0.0006 | 0.0006 | 0.0000 | 0.0000 | 0.0003 |
| MAE | 0.0207 | 0.0208 | 0.0033 | 0.0033 | 0.0120 |
| RMSE | 0.0257 | 0.0258 | 0.0041 | 0.0042 | 0.0151 |

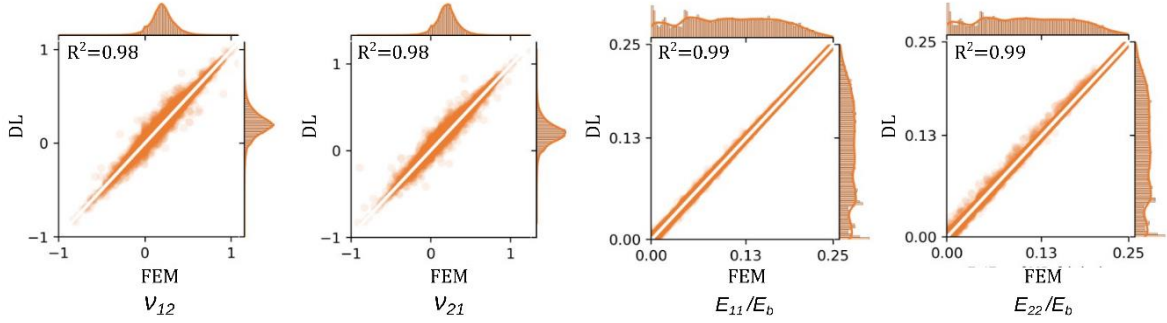

**Supplementary Figure 10.** The error plot and $R^2$ values for the FEM validation of the DL-predicted elastic properties presented in Supplementary Figure 9d.

**Supplementary Table 10.** The target, forward predicted, and numerically simulated elastic properties of the generated samples displayed in Fig. 2.

| Sample | Target | | | | Forward Predicted | | | | FEM simulated | | | |
|---|---|---|---|---|---|---|---|---|---|---|---|---|
| | $v_{12}[-]$ | $v_{21}[-]$ | $\frac{E_{11}}{E_b}[-]$ | $\frac{E_{22}}{E_b}[-]$ | $v_{12}[-]$ | $v_{21}[-]$ | $\frac{E_{11}}{E_b}[-]$ | $\frac{E_{22}}{E_b}[-]$ | $v_{12}[-]$ | $v_{21}[-]$ | $\frac{E_{11}}{E_b}[-]$ | $\frac{E_{22}}{E_b}[-]$ |
| I | −0.2 | −1.0 | 0.03 | 0.03 | −0.2 | −1.0 | 0.02 | 0.03 | −0.2 | −1.1 | 0.02 | 0.03 |
| II | −0.7 | −0.7 | 0.05 | 0.05 | −0.7 | −0.6 | 0.05 | 0.04 | −0.7 | −0.7 | 0.04 | 0.04 |
| III | −1.0 | −0.2 | 0.07 | 0.07 | −1.0 | −0.2 | 0.07 | 0.06 | −0.9 | −0.2 | 0.07 | 0.07 |
| IV | −0.5 | −0.5 | 0.08 | 0.08 | −0.5 | −0.5 | 0.08 | 0.09 | −0.5 | −0.5 | 0.09 | 0.09 |

**Supplementary Table 11.** The search space of the hyperparameter tuning pipeline for the size-agnostic model.

| Step | Process | Search Space | Selected |
|---|---|---|---|
| Data analysis | Rescaling/transforming | MinMaxScaler | |
| Model Tuning | Hidden layers | $1-7$ | 6 |
| | Neurons $1^{st}$ hidden layer | 128,256,512,1024 | 512 |
| | Network shape | Triangle | |
| | Activation function | ReLU | |
| | Last layer activation function | ReLU | |
| | Optimizer | Adam | |
| | Step size | 0.0001 | |
| | Loss function | MSE | |
| | Batch normalization | No | |
| | Targets | 4 | |
| Fitting | Batch size | $32, 64, 128, 256$ | 128 |
| | Epochs | 200 | |
| Cross-validation | K-folds | 3 | |
| | Data split | $60 - 30 - 10$ | |
| | Scoring | MSE, $R^2$ | |

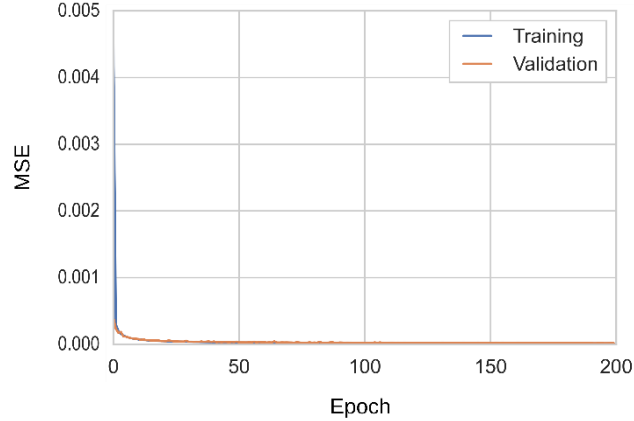

**Supplementary Figure 11.** The MSE graph for the size-agnostic model.

**Supplementary Table 12.** Regression metrics. The evaluation of the size-agnostic model using the test dataset.

| Index | $v_{12}$ | $v_{21}$ | $E_{11}$ | $E_{22}$ | Overall |
|---|---|---|---|---|---|
| $R^2$ | 0.9911 | 0.9915 | 0.9996 | 0.9995 | 0.9954 |
| MSE | 0.0005 | 0.0004 | 0.0000 | 0.0000 | 0.0002 |
| MAE | 0.0151 | 0.0144 | 0.0005 | 0.0005 | 0.0076 |
| RMSE | 0.0214 | 0.0208 | 0.0007 | 0.0007 | 0.0109 |

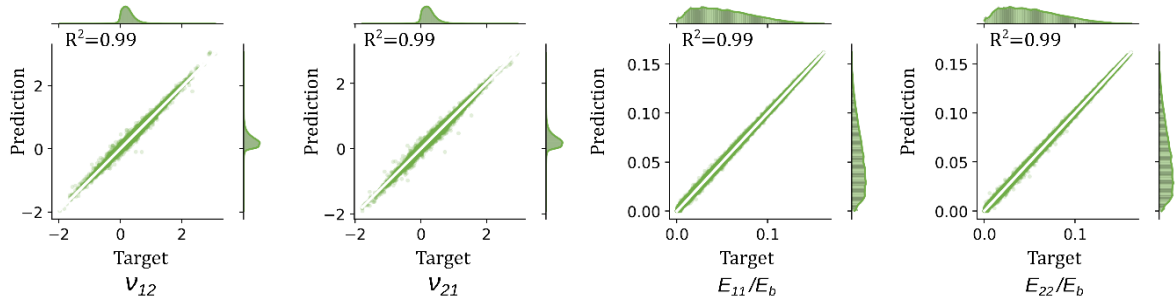

**Supplementary Figure 12.** The evaluation of the size agnostic model. The prediction *vs.* target values and the coefficients of determination for the test datasets are presented.

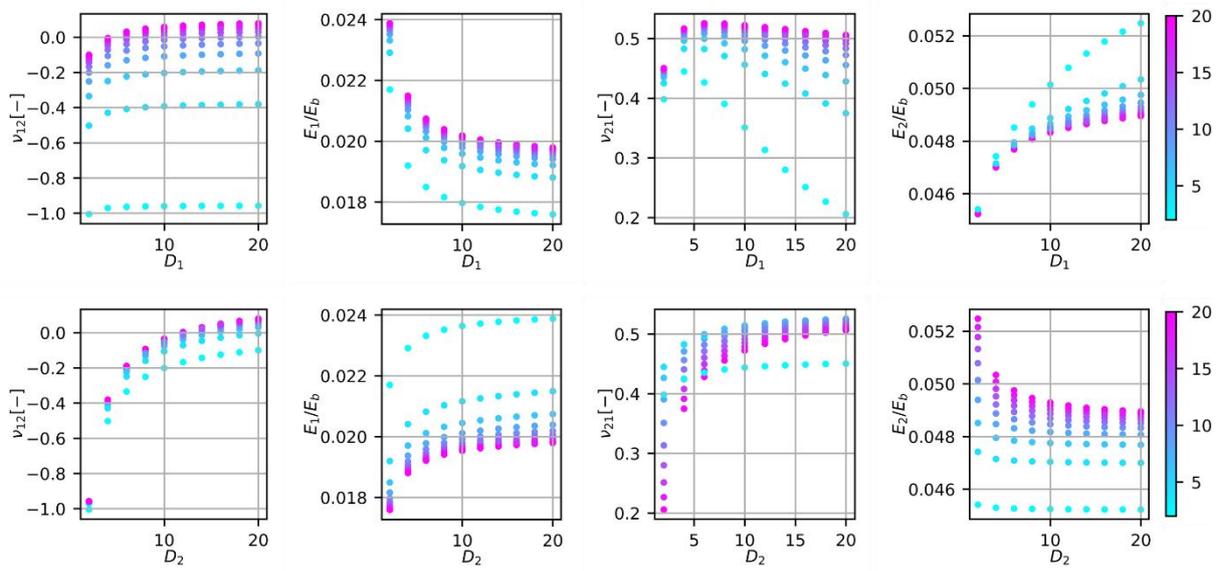

**Supplementary Figure 13.** 2D plots showing the evolution of the elastic properties based on changes in $D_1$ and $D_2$ for a representative case.